\documentclass[12pt,a4paper]{article}
\usepackage[latin1]{inputenc}
\usepackage{amsmath}
\usepackage{amsfonts}
\usepackage{amssymb}
\usepackage[left=2cm,right=2cm,top=2cm,bottom=2cm]{geometry}
\usepackage{tabularx}
\usepackage{tabularx}
\usepackage{sidecap}
\usepackage{setspace}
\usepackage{graphicx}
\usepackage{cite}
\RequirePackage[colorlinks,citecolor=blue,urlcolor=blue,linkcolor=blue]{hyperref}
\begin{document}
\title{\begin{small}
Published in \textbf{Nucl.Phys. B877 (2013) 321-342 }\,\textcolor{blue}{10.1016/j.nuclphysb.2013.10.011}
\end{small} \textbf{Quasi-degenerate Neutrino mass models and their significance: A model independent investigation}}
\author{S.~Roy\footnote{meetsubhankar@gmail.com} and N.~N.~Singh\footnote{nimai03@yahoo.com}\\
Department of Physics, Gauhati University, Guwahati, Assam-781014, India}
\date{}
\maketitle

\begin{abstract}
The prediction of possible ordering of neutrino masses relies mostly on the model selected. Alienating the $\mu-\tau$ interchange symmetry from discrete flavour symmetry based models, turns the neutrino mass matrix less predictive. But this inspires one to seek the answer from other phenomenological frameworks. We need a proper parametrization of the neutrino mass matrices concerning individual hierarchies. In the present work, we attempt to study the six different cases of Quasi-degenerate (QDN) neutrino models. The related mass matrices, $m_{LL}^{\nu}$ are parametrized with two free parameters ($\alpha,\eta$) and standard Wolfenstein parameter ($\lambda$). The input mass scale $m_0$ is selected around $\sim 0.08\,eV$. We begin with a $\mu-\tau$ symmetric neutrino mass matrix tailed by a correction from charged lepton sector. The parametrization accentuates the existence of four independent texture zero building block matrices which are common to all the QDN  models under $\mu-\tau$ symmetric framework. These remain invariant  irrespective of any choice of solar angle. In our parametrization, the neutrino sector controls the solar angle, whereas the reactor and atmospheric angles are dictated by the charged lepton sector. In the framework of oscillation experiments, cosmological observation  and future experiments involving $\beta$-decay and $0\nu\beta\beta$ experiments, all QDN models are tested  and a reason to rule out anyone out of the six models is unfounded. A strong preference for $\sin^2\theta_{12}=0.32$ is observed for QDNH-$Type A$ model. 
\end{abstract}

\section{Introduction}

One of the most challenging riddles of neutrino physics is to trace out the exact ordering of the absolute neutrino masses. The Quasi degenerate hierarchy \cite{Fukugita:1998vn,Carone:1997bb,Caldwell:1993kn, Lipmanov:2003mq, Joshipura:2000sf,Petcov:1993rk,Joshipura:2010qx,Antusch:2004xd,Binetruy:1996cs,Branco:1998bw,
Branco:1998hm,Francis:2012jj, Francis:2012jk,Singh:2009kf} among all the three possibilities, refers to the scenario when the three mass eigenvalues are of similar order, $m_1\sim m_2\sim m_3$. As the solar mass squared difference ($\Delta m_{21}^2$) is positive and the the sign of atmospheric mass squared difference ($\Delta m_{31}^2$) is unspecified, we encounter two divisions of QDN patterns: they are,
\begin{itemize}
\item `` Quasi-degenerate Normal Hierarchy (QDNH) type'' : $m_1\lesssim m_2\lesssim  m_3$,
\item ``Quasi-degenerate Inverted Hierarchy type'' (QDIH): $m_3\lesssim m_1\lesssim m_2$.
\end{itemize}
Besides, the remaining possibilities are strict ``Normal hierarchy'' (NH): $m_1<<m_2 << m_3,\, m_1\sim 0$ and ``Inverted hierarchy'' (IH): $m_3<<m_1 << m_2 ,\, m_3\sim 0$. The two Majorana phases ($\alpha,\beta$) are admitted to the diagonalized neutrino mass matrix $m_{LL}^d$, where, $m_{LL}^d=diag (m_1, m_2\,e^{i\alpha}, m_3\,e^{i\beta})$ \cite{Altarelli:1999gu}. On adopting the CP conserving cases, three subclasses corresponding to each model is generated. The CP parity patterns of the sub classes are : 
\begin{itemize}
\item Type IA: $m_{LL}^d=diag\,(+ m_1,-m_2,+m_3)$, 
\item Type IB: $m_{LL}^d=diag\,(+ m_1,+m_2,+m_3)$ and, 
\item Type IC: $m_{LL}^d=diag\,(+ m_1,+m_2,-m_3)$.
\end{itemize}

The QDN model were very often forsaken \cite{Pascoli:2007qh, Komatsu:2008hk} in view of the  neutrino-less double $\beta$ decay experiments and cosmological data. The range of absolute neutrino mass scale, $m_0$ was chosen as, $0.1\, eV-0.4\,eV$ \cite{Varzielas:2008jm} in earlier QDN models. But, the Cosmological data in concern with the sum of the three absolute neutrino masses, $\Sigma |m_i|\leq 0.28 \,eV$ \cite{Thomas:2009ae}, strongly abandons any possibility of quasi-degenerate neutrinos to exist with absolute mass scale more than $0.1\, eV$. The $\Sigma m_{i}$ corresponding to strict NH and IH scenarios are approximately $0.06\, eV$ and $0.1\, eV$ respectively. Hence the validity of both the models are beyond dispute. In the context of cosmological observation on $\Sigma\,m_{i}$ and the future experiments, we shall try to look into the possibilities related to the reanimation of the QDN models, with comparatively lower mass scale, $m_0\lesssim 0.1\, eV$.

Regarding the three unknown absolute masses, only two relations involving $m^2_{i=1,2,3}$ are known so far. In NH and IH models, this problem can be easily overcome as the lowest mass (either $m_1$ or $m_2$) is set to zero. In case of QDN model, we consider the largest mass $m_0$ as a input. Besides, there are also three mixing angles: reactor ($\theta_{13}$), solar ($\theta_{12}$) and atmospheric ($\theta_{23}$). A general neutrino mass matrix $m^{\nu}_{LL}$ carries the information of all these six quantities. For a phenomenological analysis of the QDN model, a suitable parametrization of $m^{\nu}_{LL}$ is an essential part. We shall try to design the general neutrino mass matrix, $m^{\nu}_{LL}$ with minimum numbers of free parameters. As a first approximation, $m^{\nu}_{LL}$ is assumed to follow $\mu-\tau$ symmetry \cite{Harrison:2002er, Harrison:2003aw,Ma:2004zv,Mohapatra:2006pu, He:2006qd, Plentinger:2005kx}. This symmetry keeps $\theta_{12}$ arbitrary and hence can handle both Tri-Bimaximal(TBM) mixing and deviation from it as well \cite{Francis:2012jj, Francis:2012jk,Singh:2009kf,Singh:2006uw,Singh:2006dr}. This characteristic feature of $\mu-\tau$ symmetry bears immense phenomenological importance. The expected deviations to $\theta_{13}=0$ and $\theta_{23}=\pi/4$, are controlled from charged lepton sector\cite{King:2002nf,Frampton:2004ud, Altarelli:2004jb, Antusch:2004re, Feruglio:2004gu, Mohapatra:2005yu,Antusch:2005kw,King:2005bj,Masina:2005hf,Antusch:2007rk,Duarah:2012bd}.       

We hope, this investigation on QDN mass models will serve as a platform for our future study of Baryogenesis and leptogenesis \cite{Francis:2012jj, Francis:2012jk, Singh:2009kf,Sarma:2006xk}. This investigation will require the knowledge of the texture of left handed neutrino mass matrices, $m_{LL}^{\nu}$. 

\thispagestyle{plain}
\section{Need for parametrization of a general $\mu-\tau$ symmetric mass matrix.}

The present neutrino oscillation data reports the lepton mixing angles, $\theta_{23}\sim 40^0$ and $\theta_{13}\sim 9^0$ \cite{Abe:2011fz,An:2012eh,Ahn:2012nd,Abe:2011sj,Tortola:2012te,Fogli:2012ua,
GonzalezGarcia:2012sz} which are undoubtedly deviated from what TBM mixing and BM mixing says: $\theta_{23}=45^0$ and $\theta_{13}=0^0$. A neutrino mass matrix which satisfies these properties (of BM/TBM mixing)\cite{Harrison:2002er, Harrison:2003aw,Barger:1998ta,Nomura:1998gm,Altarelli:1998nx,Harrison:2002kp,
Xing:2002sw,Harrison:2004he}, in the basis where charged lepton mass matrix, $m_{LL}^{l}$ is diagonal, $m_{LL}^l=diag(m_e,m_{\mu},m_{\tau})$, exhibits a $\mu-\tau$ interchange symmetry. With a permutation matrix, $T$ which conducts a flavor interchange $\mu \leftrightarrow\tau$, we express the texture of a $\mu-\tau$ symmetric mass matrix as in the following \cite{Grimus:2006jz, Grimus:2003wq},  

\begin{eqnarray}
\label{chap2mutau1}
T\,M_{\mu\tau}\,T=M_{\mu\tau},\Longrightarrow M_{\mu\tau}=\begin{bmatrix}
x & y & y\\
y & z & w\\
y & w & z
\end{bmatrix},
\end{eqnarray}
where,\begin{align}
T=\begin{bmatrix}
1&0&0\\
0&0&1\\
0&1&0
\end{bmatrix}.
\end{align}
We experience another form $M_{\mu\tau}$ \cite{Lam:2001fb}, different from that in Eq.(\ref{chap2mutau1}),
\begin{eqnarray}
\label{chap2mutau2}
 M_{\mu\tau}=\begin{bmatrix}
x & y & -y\\
y & z & w\\
-y & w & z
\end{bmatrix}.
\end{eqnarray}
The matrix element is invariant under the flavor interchange of $\mu \leftrightarrow -\tau$. The permutation matrix responsible for this symmetry is $T'$.
\begin{equation}
T'\,M_{\mu\tau}\,T'=M_{\mu\tau},\Longrightarrow M_{\mu\tau},
\end{equation}
 where, 
\begin{align}
T'=\begin{bmatrix}
1&0&0\\
0&0&-1\\
0&-1&0
\end{bmatrix}.
\end{align}
The presence of a $-ve$ sign before $y$ in the $1$-$3$ the element of $M_{\mu\tau}$ (see Eq.(\ref{chap2mutau2})) ensures the positivity of the mixing angles.

Except the maximal atmospheric and vanishing reactor angle, $\mu-\tau$ symmetry has no further prediction. Different discrete symmetry groups are very often combined with $\mu-\tau$ interchange symmetry in order to obtain a predictive neutrino mass matrix \cite{Grimus:2005mu}. For example, in the original Altarelli-Feruglio model \cite{Brahmachari:2008fn,Altarelli:2005yx}, the neutrino mass matrix takes the form (see Eq.(\ref{chap2mutau1})),

\begin{eqnarray}
M_{\mu\tau}=\begin{bmatrix}
a+\frac{ 2}{3} b & -\frac{b}{3} & -\frac{b}{3} \\ 
-\frac{b}{3}& \frac{2 }{3}b & a-\frac{b}{3} \\ 
-\frac{b}{3} &  a-\frac{b}{3} & \frac{2}{3}b
\end{bmatrix}, 
\end{eqnarray}
the mass eigenvalues are $m_1=a+b$, $m_2=a$ and $m_3=b-a$ which gives, $\Delta m_{sol}^2=(-b^2-2 a b)$ and $\Delta m_{atm}^2=-4 a b$. Since it is known that $\Delta m_{sol}^2 >0$, which implies $ab<0$. Hence, $a$ and $b$ must have opposite signs which in turn says, $\Delta m_{atm}^2 >0$. This model advocates for normal hierarchy of the absolute neutrino masses. In addition, it supports for TBM mixing: $\theta_{12}=\sin^{-1}(1/\sqrt{3}).$ Similarly, a neutrino mass matrix of following kind (see Eq.(\ref{chap2mutau1})),
\begin{eqnarray}
M_{\mu\tau}=\begin{bmatrix}
0 & a & a \\ 
a & b & c \\ 
a & c & b
\end{bmatrix},
\end{eqnarray}
can be related with an interesting mixing scheme called Golden ratio \cite{Kajiyama:2007gx} and dictates the mass pattern to be of normal hierarchy type.

One of the most interesting property of the $\mu-\tau$ (interchange) symmetry which is very often neglected is the arbitrariness of the solar angle $\theta_{12}$. With a proper choice of the parameters, $x,y,w$ and $z$, $\theta_{12}$ is controlled with the following relation \cite{Ma:2004zv}, intrinsic to $M_{\mu\tau}$ in Eq.(\ref{chap2mutau1}),
\begin{eqnarray}
\label{chap2tan21}
\tan 2\theta_{12}=\frac{2\sqrt{2} y}{x-w-z}.
\end{eqnarray}
For, $M_{\mu\tau}$ in Eq.(\ref{chap2mutau2}) the expression for $\tan2\theta_{12}$ is,
\begin{eqnarray}
\label{chap2tan212}
\tan 2\theta_{12}=\frac{2\sqrt{2} y}{x+w-z}
\end{eqnarray}  
It seems that $\mu-\tau$ symmetry is more natural and BM and TBM mixing schemes are certain special cases of this symmetry. In fact ,the recent result $\sin^2\theta_{12}\sim 0.32$ \cite{Tortola:2012te} deviated a little from TBM prediction ($\sin^2\theta_{12}= 0.33$), can be accommodated within the $\mu-\tau$ symmetry regime \cite{Francis:2012jj, Francis:2012jk,Singh:2009kf,Singh:2006uw,Singh:2006dr}. Neglecting the small deviations (though significant) for $\theta_{23}$ and $\theta_{13}$, we first approximate, $m_{LL}^{\nu}=M_{\mu\tau}$ and the charged lepton diagonalizing matrix, $U_{eL}=I$.  

This is an undeniable fact that the predictions of a suitable order of absolute neutrino masses are not unique and differs with the choice of models. Keeping aside all the models, which associate $M_{\mu\tau}$ with different discrete flavour symmetries, here we concentrate on a general parametrization of $M_{\mu\tau}$. The idea behind this decoupling is not to overlook the necessity of different symmetry groups, but to look into the subtle aspects of $\mu-\tau$ symmetry, starting from a phenomenological point of view. Here we emphasize on the facts that $\mu-\tau$ interchange symmetry is not partial to any hierarchy of absolute neutrino masses and has a good control over the solar angle. In the Refs.\,\cite{Roy:2012qm,Roy:2012nb}, we put forward another possible way to parametrize the neutrino mass matrix based on $\mu$-$\tau$ symmetry.

In the present article, concerning the parametrization of the $\mu-\tau$ symmetric mass matrices for different hierarchical cases, we shall stick to the second convention (see Eq.(\ref{chap2mutau2})). 
\section{Invariant building blocks of $\mu-\tau$ symmetric mass matrix}
We want to draw attention on the general texture of $M_{\mu\tau}$s satisfying BM \cite{Barger:1998ta,Nomura:1998gm,Altarelli:1998nx} and TBM \cite{Harrison:2002er, Harrison:2003aw, Harrison:2002kp,Xing:2002sw, Harrison:2004he} mixing schemes.    
\begin{align}
&M_{\mu\tau}^{BM}=\begin{bmatrix}
x & y & y \\ 
y & z & x-z \\ 
y & x-z & z
\end{bmatrix}&,\\
& M_{\mu\tau}^{TBM}=\begin{bmatrix}
x & y & y \\ 
y & z & x+y-z \\ 
y & x+y-z & z
\end{bmatrix}& .
\end{align}
It is to be noted that the above two forms of $M_{\mu\tau}$'s are in accordance with the first convention (see Eq.(\ref{chap2mutau1})). In terms of the three parameters $x,y$ and $z$, the respective mass matrices can be decomposed with certain building block matrices $I_{x},\,I^{BM,TBM}_{y}$ and $I_{z}$ in the following way.
\begin{eqnarray}
\label{chap2BMTBM1}
M^{BM}_{\mu\tau}&= & x I_x + y I^{BM}_y + z I_z,\\
\label{chap2BMTBM2}
M^{TBM}_{\mu\tau}&= & x I_x + y I^{TBM}_y + z I_z. 
\end{eqnarray}
Where,
\begin{align}
I_{x}=\begin{bmatrix}
1 & 0 & 0 \\ 
0 & 0 & 1 \\ 
0 & 1 & 0
\end{bmatrix},\, I_{z}=\begin{bmatrix}
0 & 0 & 0 \\ 
0  & 1 & -1 \\ 
0  & -1 &  1
\end{bmatrix},\\
I_{y}^{BM}=\begin{bmatrix}
0 & 1 & 1 \\ 
1 & 0 & 0 \\ 
1  & 0  & 0
\end{bmatrix}, \, I_{y}^{TBM}=\begin{bmatrix}
0 & 1 & 1 \\ 
1 & 0 & 1 \\ 
1 & 1 & 0
\end{bmatrix}.
\end{align}
There is a distinct change in the texture of $I_{y}$, as the mixing pattern transits from BM to TBM. $I_y^{BM}$ and $I_{y}^{TBM}$ have the diagonalizing matrices, $U_{BM}= R_{23}(\theta_{23}=-\pi/4). R_{13}(\theta_{13}=0).R_{12}(\theta_{12}=-\pi/4)$ and $U_{TBM}= R_{23}(\theta_{23}=-\pi/4). R_{13}(\theta_{13}=0).R_{12}(\theta_{12}=\sin^{-1}(1/\sqrt{3}))$ respectively and thus carry the signatures of respective models. For, $I_{x,z}$ the diagonalizing matrices are, $U_{x,y}=R_{23}(\theta_{23}=-\pi/4)$ . 

We insist on the possibility of finding out certain building blocks of $M_{\mu\tau}$ that will remain invariant at the face of any mixing schemes (BM or TBM) or simply independent of any $\theta_{12}$ in general. With this idea, four such independent texture-zero matrices, $I_{i=0,1,2,3}$ are posited (see Table.(\ref{chap2building})). On considering the fact that a general $M_{\mu\tau}$ is capable of holding four free parameters at the most (if $\alpha$ and $\beta$ are specified ), we parametrize $M_{\mu\tau}$ for QDNH-$Type\, IA$ case in the following way. 
\begin{align}
\label{chap2QNHA}
\quad M_{\mu\tau}=& I_0-(\beta- \frac{\alpha}{2})I_1 +  2\alpha (\eta^2-\frac{1}{4})I_2 +\alpha \eta (1-2\eta^2)^{1/2} I_{3}. \\
 =& \begin{pmatrix}
\alpha-\beta-2 \alpha \eta^2 & -\alpha \eta (1-2 \eta^2)^{\frac{1}{2}} & \alpha \eta (1-2 \eta^2)^{\frac{1}{2}} \\ 
-\alpha \eta (1-2 \eta^2)^{\frac{1}{2}} & \frac{1}{2}-\frac{\beta}{2}+\alpha \eta^2 & \frac{1}{2} +\frac{\beta}{2} -\alpha \eta^2 &\\ 
\alpha \eta (1-2 \eta^2)^{\frac{1}{2}} & \frac{1}{2} +\frac{\beta}{2} -\alpha \eta^2& \frac{1}{2}-\frac{\beta}{2}+\alpha \eta^2 &
\end{pmatrix}
\end{align}  
Here, $\alpha$, $\beta$ and $\eta$ are three free parameters and the mass matrix  is normalized with input parameter $m_0$. The parameters, $\alpha$ and $\beta$ are related with absolute masses of three neutrinos. The quantity, $m_{0}$ signifies the largest neutrino mass. It can be seen that whatever may be the changes in mixing schemes, the basic building blocks are not affected. The free parameter $\eta$ dictates the solar angle. $\eta=1/2,\,1/\sqrt{6}$, correspond to BM and TBM mixing respectively. In contrast to $M_{\mu\tau}$s in Eqs.\,(\ref{chap2BMTBM1})-(\ref{chap2BMTBM2}), the corresponding mass matrices are,
\begin{eqnarray}
 M_{\mu\tau}^{BM} = I_0-(\beta-\frac{\alpha}{2})I_1 + 0 I_2 + \frac{1}{2\sqrt{2}}\alpha I_{3},\\
 M_{\mu\tau}^{TBM}= I_0-(\beta-\frac{\alpha}{2})I_1 -\frac{1}{6}\alpha I_2 + \frac{1}{3}\alpha I_{3}. 
\end{eqnarray}
Here we want to add that with $\eta=2/5$, $\sin^{2}\theta_{12}=0.32$ (best-fit)\cite{Tortola:2012te} can be obtained.  
\begin{table}
\setlength{\tabcolsep}{1.6 em}
\centering
\begin{tabular}{l c c c}
\hline 
\hline
$I_{i}$ & &$I_{i}^{diag}$  & $U_{i}$ \\ 
\hline
\hline
\\ 
$I_0$ & $\frac{1}{2}\begin{bmatrix}
0 & 0 &0\\
0 & 1  &1\\
0 & 1  &1
\end{bmatrix}$ & $ \begin{bmatrix}
0 & 0 &0\\
0 & 0  &0\\
0 & 0  &1
\end{bmatrix}$ &$\begin{bmatrix}
1 & 0 & 0 \\ 
0 &\frac{1}{\sqrt{2}} & \frac{1}{\sqrt{2}}\\ 
0 & -\frac{1}{\sqrt{2}}& \frac{1}{\sqrt{2}}
\end{bmatrix}$ \\ 
 
$I_1$ & $\frac{1}{2}\begin{bmatrix}
2 & 0 &0\\
0 & 1  &-1\\
0 &  -1  &1
\end{bmatrix}$ & $ \begin{bmatrix}
1 & 0 &0\\
0 & 1  &0\\
0 & 0  &0
\end{bmatrix}$& $\begin{bmatrix}
1 & 0 & 0 \\ 
0 &\frac{1}{\sqrt{2}} & \frac{1}{\sqrt{2}}\\ 
0 & -\frac{1}{\sqrt{2}}& \frac{1}{\sqrt{2}}
\end{bmatrix}$\\ 
 
$I_2$ & $
\frac{1}{2}\begin{bmatrix}
-2 & 0   & 0 \\ 
 0 &  1   & -1 \\ 
 0 &  -1   & 1
\end{bmatrix}$ &$ \begin{bmatrix}
-1 & 0 &0\\
0 & 1  &0\\
0 & 0  &0
\end{bmatrix}$ & $\begin{bmatrix}
1 & 0 & 0 \\ 
0 &\frac{1}{\sqrt{2}} & \frac{1}{\sqrt{2}}\\ 
0 & -\frac{1}{\sqrt{2}}& \frac{1}{\sqrt{2}}
\end{bmatrix}$\\ 

$I_3$ & $\begin{bmatrix}
0 &   -1 &  1 \\ 
-1  &   0 &  0 \\ 
1  &   0   &  0
\end{bmatrix}$ &$ \begin{bmatrix}
-\sqrt{2} & 0 &0\\
0 & \sqrt{2}  &1\\
0 & 1  &0
\end{bmatrix} $ & $\begin{bmatrix}
\frac{1}{\sqrt{2}} & \frac{1}{\sqrt{2}} & 0\\
-\frac{1}{2} & \frac{1}{2} & \frac{1}{\sqrt{2}}\\
\frac{1}{2} & -\frac{1}{2} & \frac{1}{\sqrt{2}}
\end{bmatrix}$\\ 
\\
\hline 
\end{tabular}
\caption{\footnotesize The texture of the invariant building blocks $I_{i=0,1,2,3}$, the diagonalized blocks $I^{diag}_{i=0,1,2,3}$ and the corresponding diagonalizing matrices ($U_i$).} 
\label{chap2building}
\end{table} 
It can be seen that, $I_0 +I_1=I$, the identity matrix. Also, from Table.(\ref{chap2building}), this is interesting to note that the diagonalizing matrix of $I_{3}$ is none other than $U_{BM}$. 

There are certain significant features of this parametrization. With same building block matrices, we can extend the parametrization of $M_{\mu\tau}$s for other five QDN and even for the NH and IH cases also. For example, similar to Eq. (\ref{chap2QNHA}), a rearrangement of the free parameters $(\alpha,\beta,\eta)$, and $I_i$s we parametrize $M_{\mu\tau}$ for QDIH-$Type\, IA$ case in the following way.
\begin{align}
\label{chap2QDIA}
M_{\mu\tau} = \beta I_0-\left(1-\frac{\alpha}{2}\right)I_1 + 2\alpha \left(\eta^2-\frac{1}{4}\right)I_2 +\alpha \eta \left(1-2\eta^2\right)^{1/2} I_{3}.
\end{align}
Upon considering $\beta=\alpha$, in Eq.(\ref{chap2QNHA}), we get $M_{\mu\tau}$ satisfying strict NH-$Type\,IA$ condition.
\begin{align}
M_{\mu\tau}=I_0-\frac{\alpha}{2}I_1 & +2\alpha \left(\eta^2-\frac{1}{4}\right)I_2+\alpha \eta \left(1-2\eta^2\right)^{1/2} I_{3}.
\end{align}
Similarly, with $\beta=0$, in Eq.(\ref{chap2QDIA}), we obtain a $M_{\mu\tau}$ that represents IH-$Type\,IA$ case. 
\begin{align}
M_{\mu\tau} = 0 I_0-\left(1-\frac{\alpha}{2}\right)I_1 & + 2\alpha \left(\eta^2-\frac{1}{4}\right)I_2+\alpha \eta \left(1-2\eta^2\right)^{1/2} I_{3}.
\end{align}
Similarly, we can formulate the same for other cases also. The details are shown in Table.(\ref{chap2table2}) and Table.(\ref{chap2table3})

In this present approach, the mass parameters and the mixing angle parameters are decoupled. A single expression of $\tan2\theta_{12}$ for all the eleven cases is,
\begin{eqnarray}
\label{chap2tan2}
\tan 2\theta_{12}&=&\frac{2\sqrt{2}\eta (1-2\eta^2)^{1/2}}{1-4\eta^2},\\
\text{or,}\quad\sin^{2}\theta_{12}&=&2\eta^2. 
\end{eqnarray}.
\begin{table}
\centering
\begin{tabular}{l c c}
\hline
\hline
 QDN-NH,IH     &    $M_{\mu\tau}(\alpha,\beta,\eta)/m_0$ & $m_i/m_0$ \\
\hline
\hline\\
QDNH-$IA$ :  & \begin{tabular}{c}
$\begin{bmatrix}
\alpha-\beta-2 \alpha \eta^2 & -\alpha \eta (1-2 \eta^2)^{\frac{1}{2}} & \alpha \eta (1-2 \eta^2)^{\frac{1}{2}} \\ 
-\alpha \eta (1-2 \eta^2)^{\frac{1}{2}} & \frac{1}{2}-\frac{\beta}{2}+\alpha \eta^2 & \frac{1}{2} +\frac{\beta}{2} -\alpha \eta^2 &\\ 
 \alpha \eta (1-2 \eta^2)^{\frac{1}{2}} & \frac{1}{2} +\frac{\beta}{2} -\alpha \eta^2  & \frac{1}{2}-\frac{\beta}{2}+\alpha \eta^2 &
\end{bmatrix}$ \\ 
 \\
\begin{footnotesize}
$=I_0-(\beta- \frac{\alpha}{2})I_1 +  2\alpha (\eta^2-\frac{1}{4})I_2
 +\alpha \eta (1-2\eta^2)^{1/2} I_{3}.$
\end{footnotesize} \\ 
\end{tabular}   &  \begin{tabular}{c}
$\alpha-\beta$ \\ 

$-\beta$ \\ 
 
$1$ \\ 
\end{tabular}\\ 
\\
QDNH-$IB$ :  & \begin{tabular}{c}
$\begin{bmatrix}
\beta+2 \alpha \eta^2 -\alpha& \alpha \eta (1-2 \eta^2)^{\frac{1}{2}} & -\alpha \eta (1-2 \eta^2)^{\frac{1}{2}} \\ 
\alpha \eta (1-2 \eta^2)^{\frac{1}{2}} & \frac{1}{2}+\frac{\beta}{2}-\alpha \eta^2 & \frac{1}{2} -\frac{\beta}{2} +\alpha \eta^2 \\ 
-\alpha \eta (1-2 \eta^2)^{\frac{1}{2}} & \frac{1}{2} -\frac{\beta}{2} +\alpha \eta^2& \frac{1}{2}+\frac{\beta}{2}-\alpha \eta^2 &
\end{bmatrix}$ \\ 
 \\
\begin{footnotesize}
$ =I_0+(\beta-\frac{\alpha}{2})I_1 -2\alpha (\eta^2-\frac{1}{4})I_2 -\alpha \eta (1-2\eta^2)^{1/2} I_{3}$
\end{footnotesize} \\ 
\end{tabular}   &  \begin{tabular}{c}
$\beta-\alpha$ \\ 

$\beta$ \\ 
 
$1$ \\ 
\end{tabular}\\
\\
QDNH-$IC$ :  & \begin{tabular}{c}
$\begin{bmatrix}
\beta+2 \alpha \eta^2-\alpha & \alpha \eta (1-2 \eta^2)^{\frac{1}{2}} & -\alpha \eta (1-2 \eta^2)^{\frac{1}{2}} \\ 
\alpha \eta (1-2 \eta^2)^{\frac{1}{2}} & \frac{\beta}{2}-\alpha \eta^2-\frac{1}{2} & \alpha \eta^2-\frac{1}{2} -\frac{\beta}{2} \\ 
-\alpha \eta (1-2 \eta^2)^{\frac{1}{2}} &\alpha \eta^2-\frac{1}{2} -\frac{\beta}{2}  & \frac{\beta}{2}-\alpha \eta^2-\frac{1}{2} &
\end{bmatrix}$ \\ 
 \\
\begin{footnotesize}
$ =-I_0+(\beta-\frac{\alpha}{2})I_1 -2\alpha (\eta^2-\frac{1}{4})I_2 -\alpha \eta (1-2\eta^2)^{1/2} I_{3}$
\end{footnotesize} \\ 
\end{tabular}   &  \begin{tabular}{c}
$\beta-\alpha$ \\ 

$\beta$ \\ 
 
$-1$ \\ 
\end{tabular}\\

\\
\\
QDIH-$IA$ :  & \begin{tabular}{c}
$\begin{bmatrix}
 \alpha -2 \alpha  \eta ^2 -1 & -\alpha  \eta  (1-2 \eta ^2)^{\frac{1}{2}} & \alpha  \eta  (1-2 \eta ^2)^{\frac{1}{2}} \\
 -\alpha  \eta  (1-2 \eta ^2)^{\frac{1}{2}} & \frac{\beta}{2}+\alpha  \eta ^2-\frac{1}{2} & \frac{1}{2}+\frac{\beta}{2}-\alpha  \eta ^2 \\
\alpha  \eta  (1-2 \eta ^2)^{\frac{1}{2}} & \frac{1}{2}+\frac{\beta}{2}-\alpha  \eta ^2 &\frac{\beta}{2}+\alpha  \eta ^2 -\frac{1}{2}
\end{bmatrix}$ \\ 
 \\
\begin{footnotesize}
$ =\beta I_0-(1-\frac{\alpha}{2})I_1 + 2\alpha (\eta^2-\frac{1}{4})I_2 +\alpha \eta (1-2\eta^2)^{1/2} I_{3}.$
\end{footnotesize} \\ 
\end{tabular}   &  \begin{tabular}{c}
$\alpha-1$ \\ 

$-1$ \\ 
 
$\beta$ \\ 
\end{tabular}\\
\\
\\
QDIH-$IB$ :  & \begin{tabular}{c}
$\left[
\begin{array}{ccc}
 1-\alpha +2 \alpha  \eta ^2 & \alpha  \eta  (1-2 \eta ^2)^{\frac{1}{2}} & -\alpha  \eta  (1-2 \eta ^2)^{\frac{1}{2}} \\
\alpha  \eta  (1-2 \eta ^2)^{\frac{1}{2}}  & \frac{1}{2}+\frac{\beta}{2}-\alpha  \eta ^2 & \frac{\beta}{2}+\alpha  \eta ^2 -\frac{1}{2}\\
-\alpha  \eta  (1-2 \eta ^2)^{\frac{1}{2}} & \frac{\beta}{2}+\alpha  \eta ^2 -\frac{1}{2} & \frac{1}{2}+\frac{\beta} {2} -\alpha  \eta ^2
\end{array}
\right]$ \\ 
 \\
\begin{footnotesize}
$ =\beta I_0+(1-\frac{\alpha}{2})I_1 -2\alpha (\eta^2-\frac{1}{4})I_2 -\alpha \eta (1-2\eta^2)^{1/2} I_{3}.$
\end{footnotesize} \\ 
\end{tabular}   &  \begin{tabular}{c}
$1-\alpha$ \\ 

$1$ \\ 
 
$\beta$ \\ 
\end{tabular}\\
\\
QDIH-$IC$ :  & \begin{tabular}{c}
$\left[
\begin{array}{ccc}
 1-\alpha +2 \alpha  \eta ^2 & \alpha  \eta  (1-2 \eta ^2)^{\frac{1}{2}} & -\alpha  \eta (1-2 \eta ^2)^{\frac{1}{2}}\\
\alpha  \eta  (1-2 \eta ^2)^{\frac{1}{2}} & \frac{1}{2}-\frac{\beta}{2}-\alpha  \eta ^2 & \alpha  \eta ^2-\frac{\beta}{2}-\frac{1}{2} \\
 -\alpha  \eta (1-2 \eta ^2)^{\frac{1}{2}} & \alpha  \eta ^2-\frac{\beta}{2}-\frac{1}{2}  & \frac{1}{2}-\frac{\beta}{2}-\alpha  \eta ^2
\end{array}
\right]$ \\ 
 \\
\begin{footnotesize}
$ = -\beta I_0+(1-\frac{\alpha}{2})I_1 -2\alpha (\eta^2-\frac{1}{4})I_2 -\alpha \eta (1-2\eta^2)^{1/2} I_{3}.$
\end{footnotesize} \\ 
\end{tabular}   &  \begin{tabular}{c}
$1-\alpha$ \\ 

$1$ \\ 
 
$-\beta$ \\ 
\end{tabular}\\
\\
\hline 
\end{tabular}
\caption{\footnotesize The parametrization of $M_{\mu\tau}$ for six different QDN cases with three free parameters $(\alpha,\beta,\eta)$ with four basic building blocks $I_{i=0,1,2,3}$. $m_0$ is the input parameter.}
\label{chap2table2}
\end{table}
\begin{table}
\centering
\begin{tabular}{l c c}
\hline
\hline
 NH,IH     &    $M_{\mu\tau}(\alpha,\eta)/m_0$ & $m_i/m_0$ \\
\hline
\hline\\
NH-$IA$ :  & \begin{tabular}{c}
$\left[
\begin{array}{ccc}
 -2 \alpha  \eta ^2 & -\alpha  \eta (1-2 \eta ^2)^{\frac{1}{2}} & \alpha  \eta (1-2 \eta ^2)^{\frac{1}{2}} \\
-\alpha  \eta (1-2 \eta ^2)^{\frac{1}{2}} & \frac{1}{2}-\frac{\alpha}{2} +\alpha  \eta ^2 & \frac{1}{2}+ \frac{\alpha}{2} -\alpha  \eta ^2 \\
\alpha  \eta (1-2 \eta ^2)^{\frac{1}{2}}&    \frac{1}{2}+ \frac{\alpha}{2} -\alpha  \eta ^2 				& \frac{1}{2}-\frac{\alpha}{2} +\alpha  \eta ^2
\end{array}
\right]$ \\ 
 \\
\begin{footnotesize}
$=I_0-\frac{\alpha}{2}I_1 +  2\alpha (\eta^2-\frac{1}{4})I_2
 +\alpha \eta (1-2\eta^2)^{1/2} I_{3}.$
\end{footnotesize} \\ 
\end{tabular}   &  \begin{tabular}{c}
$0$ \\ 

$-\alpha$ \\ 
 
$1$ \\ 
\end{tabular}\\
\\
NH-$IB$ :  & \begin{tabular}{c}
$\left[
\begin{array}{ccc}
 2 \alpha  \eta ^2 & \alpha  \eta (1-2 \eta ^2)^{\frac{1}{2}} & -\alpha  \eta (1-2 \eta ^2)^{\frac{1}{2}} \\
  \alpha  \eta (1-2 \eta ^2)^{\frac{1}{2}}          & \frac{1}{2}+\frac{\alpha}{2}-\alpha  \eta ^2 & \frac{1}{2}-\frac{\alpha}{2}+\alpha\eta ^2 \\
 -\alpha  \eta (1-2 \eta ^2)^{\frac{1}{2}}  &     \frac{1}{2}-\frac{\alpha}{2}+\alpha\eta ^2       & \frac{1}{2}+\frac{\alpha}{2}-\alpha  \eta ^2
 \end{array}
 \right]$ \\ 
 \\
\begin{footnotesize}
$=I_0+\frac{\alpha}{2}I_1 -2\alpha (\eta^2-\frac{1}{4})I_2 -\alpha \eta (1-2\eta^2)^{1/2} I_{3}$
\end{footnotesize} \\ 
\end{tabular}   &  \begin{tabular}{c}
$0$ \\ 

$\alpha$ \\ 
 
$1$ \\ 
\end{tabular}\\
\\
NH-$IC$ :  & \begin{tabular}{c}
$\left[
\begin{array}{ccc}
 2 \alpha  \eta ^2 & \alpha  \eta (1-2 \eta ^2)^{\frac{1}{2}} & -\alpha  \eta (1-2 \eta ^2)^{\frac{1}{2}} \\
 \alpha  \eta (1-2 \eta ^2)^{\frac{1}{2}} & \frac{\alpha}{2}-\alpha  \eta ^2-\frac{1}{2} & \alpha  \eta ^2-\frac{\alpha}{2}-\frac{1}{2} \\
 -\alpha  \eta (1-2 \eta ^2)^{\frac{1}{2}}& \alpha  \eta ^2-\frac{\alpha}{2}-\frac{1}{2}  & \frac{\alpha}{2}-\alpha  \eta ^2-\frac{1}{2}
\end{array}
\right]$ \\ 
 \\
\begin{footnotesize}
$=-I_0-\frac{\alpha}{2}I_1 -2\alpha (\eta^2-\frac{1}{4})I_2 -\alpha \eta (1-2\eta^2)^{1/2} I_{3}$.
\end{footnotesize} \\ 
\end{tabular}   &  \begin{tabular}{c}
$0$ \\ 

$\alpha$ \\ 
 
$-1$ \\ 
\end{tabular}\\
\\
IH-$IA$ :  & \begin{tabular}{c}
$\left[
\begin{array}{ccc}
 \alpha -2 \alpha  \eta ^2 -1 & -\alpha  \eta (1-2 \eta ^2)^{\frac{1}{2}} & \alpha  \eta  (1-2 \eta ^2)^{\frac{1}{2}} \\
 -\alpha  \eta (1-2 \eta ^2)^{\frac{1}{2}}  & \alpha  \eta ^2-\frac{1}{2} & \frac{1}{2}-\alpha  \eta ^2 \\
 -\alpha  \eta (1-2 \eta ^2)^{\frac{1}{2}}  & \frac{1}{2}-\alpha  \eta ^2 & \alpha  \eta ^2-\frac{1}{2}
\end{array}
 \right]$ \\ 
 \\
\begin{footnotesize}
$ =0 I_0-(1-\frac{\alpha}{2})I_1 + 2\alpha (\eta^2-\frac{1}{4})I_2 +\alpha \eta (1-2\eta^2)^{1/2} I_{3}.$
\end{footnotesize} \\ 
\end{tabular}   &  \begin{tabular}{c}
$\alpha-1$ \\ 

$-1$ \\ 
 
$0$ \\ 
\end{tabular}\\
\\
IH-$IB$ :  & \begin{tabular}{c}
$\left[
\begin{array}{ccc}
 1-\alpha + 2 \alpha  \eta ^2 & \alpha \eta (1-\eta^2)^{\frac{1}{2}}    & -\alpha \eta (1-\eta^2)^{\frac{1}{2}}\\
\alpha \eta (1-\eta^2)^{\frac{1}{2}}   & \frac{1}{2}-\alpha  \eta ^2 & \alpha  \eta ^2 -\frac{1}{2}\\
-\alpha \eta (1-\eta^2)^{\frac{1}{2}} &  \alpha  \eta ^2 -\frac{1}{2}                         & \frac{1}{2}-\alpha  \eta ^2
\end{array}
\right]$ \\ 
 \\
\begin{footnotesize}
$ =0 I_0-(1-\frac{\alpha}{2})I_1 + 2\alpha (\eta^2-\frac{1}{4})I_2 +\alpha \eta (1-2\eta^2)^{1/2} I_{3}.$
\end{footnotesize} \\ 
\end{tabular}   &  \begin{tabular}{c}
$1-\alpha$ \\ 

$1$ \\ 
 
$0$ \\ 
\end{tabular}\\
\\
\hline
\end{tabular}
\caption{The extension of the parametrization to NH and IH models. It can be seen that only two free parameters $\alpha$ and $\eta$ are required to parametrize the mass matrices.}
\label{chap2table3}
\end{table}
\section{The input parameter $m_{0}$ for QDN model}
\label{chap2m0}
\begin{figure}
\begin{center}
\includegraphics[scale=0.8]{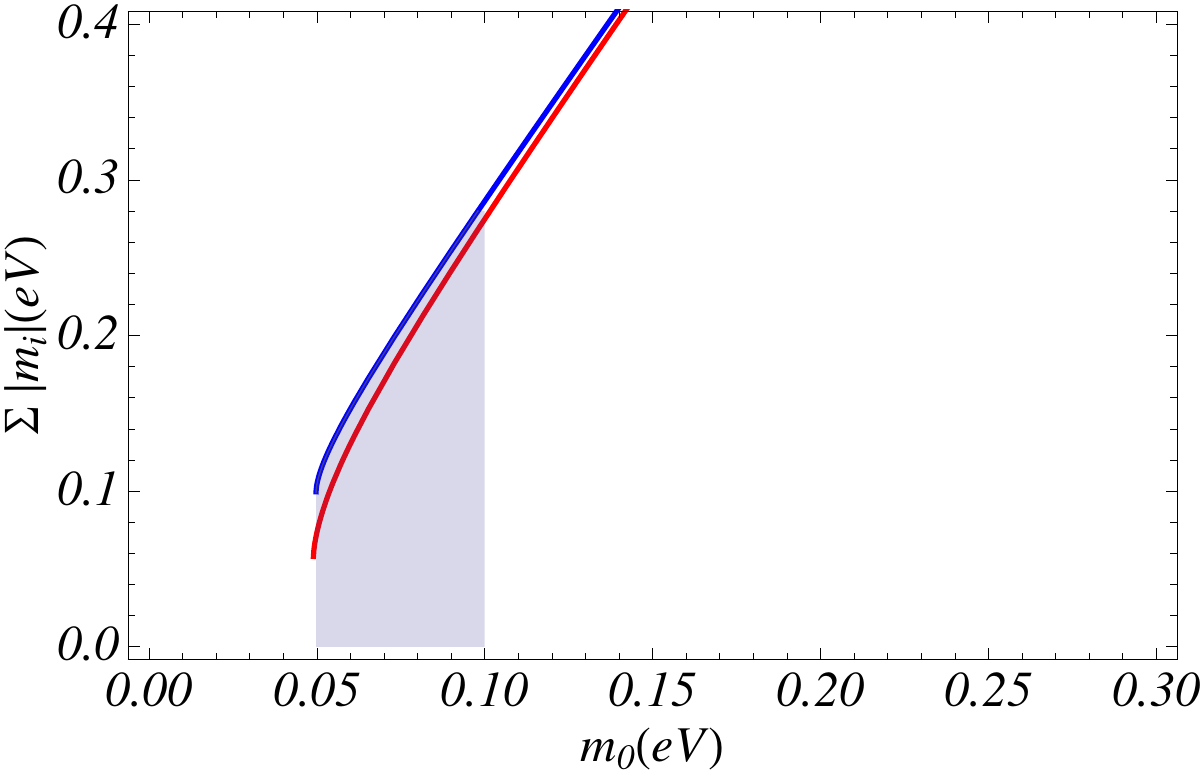} 
\caption{\footnotesize $\Sigma m_i$ vs the input parameter $m_0$. Corresponding to the cosmological upper bound $\Sigma m_{i}\lesssim 0.28\, eV$ and beyond $m_{0}> 0.05 \, eV$, $\Sigma m_i$ is imaginary, we get a range of $m_0$ as $[\,0.05,\,0.1\,]\, eV$. The Red stands for QDNH case while Blue signifies the case of QDIH}
\label{chap2fig1}
\end{center}
\end{figure}

\begin{figure}
\begin{center}
\includegraphics[scale=0.84]{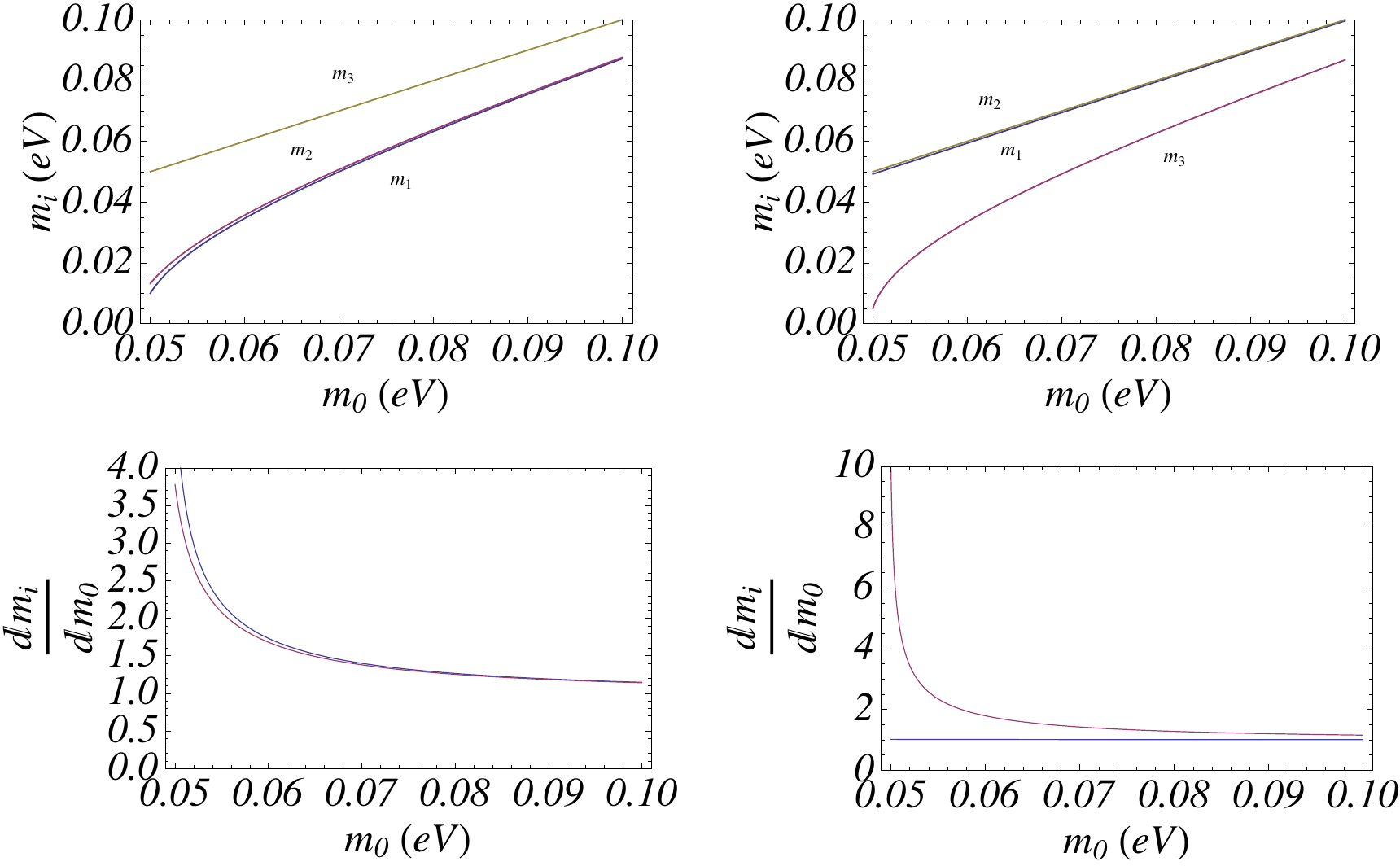} 
\caption{\footnotesize Study of $m_{i}$ vs $m_0$ (top-left: QDNH case, top-right: QDIH case) and $dm_i/dm_0$ vs $m_0$ (bottom-left: QDNH case, bottom-right: QDIH case). }
\label{chap2fig2}
\end{center}
\end{figure}

In either of the two QDN cases, $m_0$ represents the largest absolute neutrino mass. For QDNH cases, we use the following relations to work out the neutrino masses $m_i$. 
\begin{eqnarray}
m_1 &=& m_0\sqrt{1-\frac{\Delta m_{atm}^2}{m_0 ^2}},\\
m_2 &=& m_0 \sqrt{1+\frac{\Delta m_{sol}^2}{m_0 ^2}-\frac{\Delta m_{atm}^2}{m_0 ^2}},\\
m_3 &=& m_0.
\end{eqnarray} 
For, QDIH cases, we use,
\begin{eqnarray}
m_1 &=& m_0\sqrt{1-\frac{\Delta m_{sol}^2}{m_0 ^2}},\\
m_2 &=& m_0,\\
m_3 &=& m_0 \sqrt{1-\frac{\Delta m_{sol}^2}{m_0 ^2}-\frac{\Delta m_{atm}^2}{m_0 ^2}}.
\end{eqnarray}
Also, we have,
\begin{eqnarray}
\Sigma m_{i}= |m_{1}|+ |m_{2}|+ |m_{3}|.
\end{eqnarray}
The present cosmological upper bound on $\Sigma m_i$ is $0.28\,eV$ \cite{Thomas:2009ae} and the best-fit values of the mass squared differences are approximately: $\Delta m_{21}^2\sim 7.6 \times 10^{-5} \, eV^2$ , $\Delta m_{31}^2\sim 2.4 \times 10^{-3} \, eV^2$ \cite{Tortola:2012te,Fogli:2012ua,GonzalezGarcia:2012sz}. From a graphical analysis of $\Sigma\, |m_i|$ vs. $m_0$ reveals that the absolute mass scale $m_0$ must lie approximately within $0.05\,eV-0.1\,eV$ (Fig.(\ref{chap2fig1})). The upper limit of $m_0$ is the direct outcome of the cosmological upper bound \cite{Thomas:2009ae}. The lower limit arises because, when $m_0\lesssim 0.05\,eV$, $m_1$, $m_2$ for QDNH case and $m_{3}$ for QDIH case become imaginary. By studying the variation of $m_i$ and corresponding slopes ($dm_{i}/d m_0$) with respect to $m_0$ (Fig.(\ref{chap2fig2})), we expect that the level of degeneracy is better for $m_0>0.07\,eV$ and approximate the range of $m_{0}$ from $0.07-0.1\,eV$. For all numerical studies we adhere to $m_0\sim 0.08\,eV$. 


\section{Endeavor to suppress the number of free parameters in QDN models }
\label{chap2etabeta}
This is clear that only two free parameters $\alpha$ and $\eta$ are required to parametrize $M_{\mu\tau}$ for NH and IH models (see Table.(\ref{chap2table3})); whereas QDN model requires three ($\alpha,\beta,\eta$) (see Table.(\ref{chap2table2})). The rejection of one parameter for NH and IH cases is natural. But we shall try to see whether under certain logical ground we can suppress the number of free parameters for QDN model or not.

We consider the example of QDNH-Type IA case. With $m_{0}\sim 0.08\, eV$, we study the ratio $\alpha:\beta$ and $\beta:\eta$ for the $2\sigma$ and $3\sigma$ ranges of the three parameters based on the Global data analysis \cite{Fogli:2012ua}. The idea behind this approach is to detect whether there exists a simple linear correlation between the parameters or not. 

\begin{figure}
\begin{center}
\includegraphics[scale=0.7]{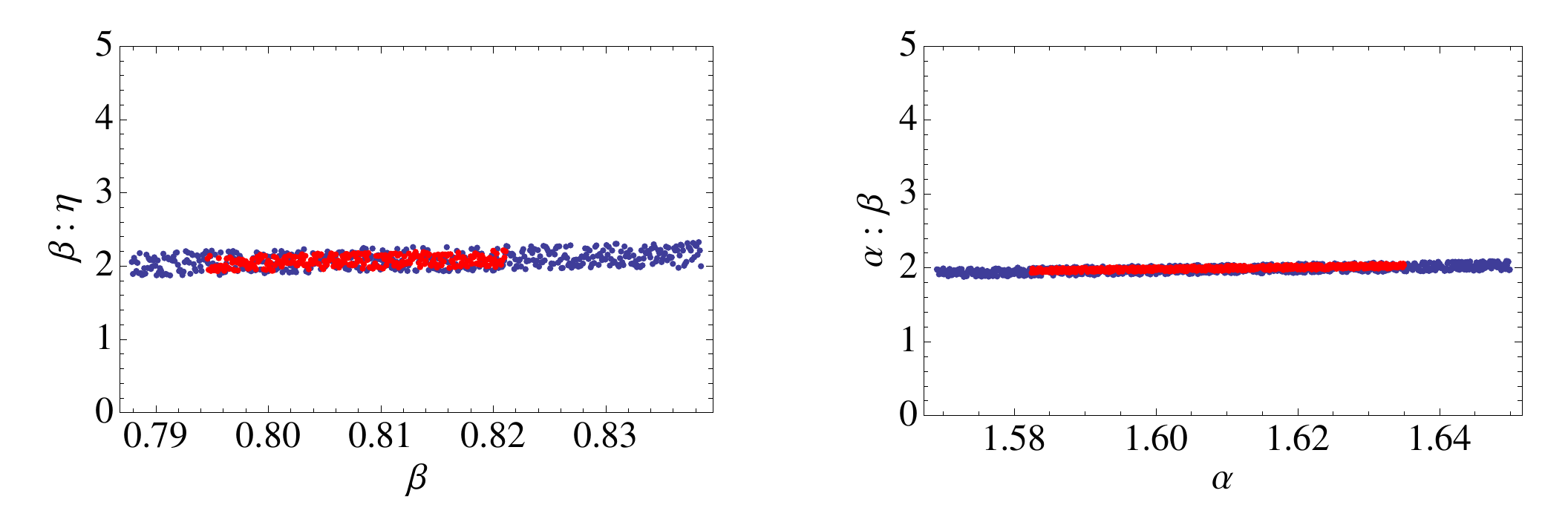}
\caption{\footnotesize In order to check the validity of the assumptions: there may lie a linear correlation between the parameters $(\alpha,\beta,\eta)$ for QDNH-IA case, we check graphically $\alpha :\beta$ and $\beta : \eta$. The analysis hints for $\beta=2\eta$ and $\alpha=2\beta$. The colour red and blue stand for the $2\sigma$ and the $3\sigma$ range respectively.}
\label{chap2fig3}
\end{center}
\end{figure}

Fig (\ref{chap2fig3})reveals such a quest is not absurd at all and we can assume, $\alpha\approxeq 2\beta$ and $\beta \approxeq 2\eta$. But the first ansatz leads to $\Delta m_{21}^2=0$ and turns out insignificant. We stick to the second ansatz. An immediate outcome is that the parameter $\eta$ which is responsible only for the mixing angle $\theta_{12}$  in the earlier parametrization of $M_{\mu\tau}(\alpha,\beta,\eta)$ is now capable of driving the mass parameters also. In other words, the arbitrariness of $\theta_{12}$ is now reduced a little. In contrast to Eq.(\ref{chap2QNHA}), for QDNH-$TypeA$ case, with normalized $m_0$, we have,
\begin{align}
\label{chap2QNHAA}
 M_{\mu\tau}(\alpha,\eta)&=I_0-(2\eta- \frac{\alpha}{2})I_1 +  2\alpha (\eta^2-\frac{1}{4})I_2+\alpha \eta (1-2\eta^2)^{1/2} I_{3},\\
&=\begin{bmatrix}
\alpha-2\eta-2 \alpha \eta^2 & -\alpha \eta (1-2 \eta^2)^{\frac{1}{2}} & \alpha \eta (1-2 \eta^2)^{\frac{1}{2}} \\ 
-\alpha \eta (1-2 \eta^2)^{\frac{1}{2}} & \frac{1}{2}-\eta+\alpha \eta^2 & \frac{1}{2} +\eta -\alpha \eta^2 \\ 
\alpha \eta (1-2 \eta^2)^{\frac{1}{2}} & \frac{1}{2} +\eta -\alpha \eta^2 & \frac{1}{2}-\eta+\alpha \eta^2 
\end{bmatrix}.
\end{align}
The ansatz $\beta=2\eta$ is applicable to other remaining QDN cases also (see Table.(\ref{chap2table4})). Needless to say, that the suppression of parameters does not affect $\tan2\theta_{12}$ in Eq.(\ref{chap2tan2}).
\begin{table}
\centering
\begin{tabular}{l c c}
\hline
\hline
 QDN-NH,IH     &    $M_{\mu\tau}(\alpha,\eta)/m_0$ & $m_i/m_0$ \\
\hline
\hline\\
QDNH-$IA$ :  & \begin{tabular}{c}
$\begin{bmatrix}
\alpha-2\eta-2 \alpha \eta^2 & -\alpha \eta (1-2 \eta^2)^{\frac{1}{2}} & \alpha \eta (1-2 \eta^2)^{\frac{1}{2}} \\ 
-\alpha \eta (1-2 \eta^2)^{\frac{1}{2}}  & \frac{1}{2}-\eta+\alpha \eta^2 & \frac{1}{2} +\eta -\alpha \eta^2 \\ 
 \alpha \eta (1-2 \eta^2)^{\frac{1}{2}}& \frac{1}{2} +\eta -\alpha \eta^2 & \frac{1}{2}-\eta+\alpha \eta^2 &
\end{bmatrix}$ \\ 
 \\
\begin{footnotesize}
$=I_0-(2\eta- \frac{\alpha}{2})I_1 +  2\alpha (\eta^2-\frac{1}{4})I_2
 +\alpha \eta (1-2\eta^2)^{1/2} I_{3}.$
\end{footnotesize} \\ 
\end{tabular}   &  \begin{tabular}{c}
$\alpha-2\eta$ \\ 

$-2\eta$ \\ 
 
$1$ \\ 
\end{tabular}\\ 
\\
QDNH-$IB$ :  & \begin{tabular}{c}
$\begin{bmatrix}
2\eta+2 \alpha \eta^2 -\alpha& \alpha \eta (1-2 \eta^2)^{\frac{1}{2}} & -\alpha \eta (1-2 \eta^2)^{\frac{1}{2}} \\ 
\alpha \eta (1-2 \eta^2)^{\frac{1}{2}} & \frac{1}{2}+\eta-\alpha \eta^2 & \frac{1}{2} -\eta +\alpha \eta^2 \\ 
-\alpha \eta (1-2 \eta^2)^{\frac{1}{2}} & \frac{1}{2} -\eta +\alpha \eta^2  & \frac{1}{2}+\eta-\alpha \eta^2 &
\end{bmatrix}$ \\ 
 \\
\begin{footnotesize}
$ =I_0+(2\eta-\frac{\alpha}{2})I_1 -2\alpha (\eta^2-\frac{1}{4})I_2 -\alpha \eta (1-2\eta^2)^{1/2} I_{3}$
\end{footnotesize} \\ 
\end{tabular}   &  \begin{tabular}{c}
$2\eta-\alpha$ \\ 

$2\eta$ \\ 
 
$1$ \\ 
\end{tabular}\\
\\
QDNH-$IC$ :  & \begin{tabular}{c}
$\begin{bmatrix}
2\eta+2 \alpha \eta^2-\alpha & \alpha \eta (1-2 \eta^2)^{\frac{1}{2}} & -\alpha \eta (1-2 \eta^2)^{\frac{1}{2}} \\ 
\alpha \eta (1-2 \eta^2)^{\frac{1}{2}}  & \eta-\alpha \eta^2-\frac{1}{2} & \alpha \eta^2-\frac{1}{2} -\eta \\ 
-\alpha \eta (1-2 \eta^2)^{\frac{1}{2}} & \alpha \eta^2-\frac{1}{2} -\eta & \eta-\alpha \eta^2-\frac{1}{2} &
\end{bmatrix}$ \\ 
 \\
\begin{footnotesize}
$ =-I_0+(2\eta-\frac{\alpha}{2})I_1 -2\alpha (\eta^2-\frac{1}{4})I_2 -\alpha \eta (1-2\eta^2)^{1/2} I_{3}$
\end{footnotesize} \\ 
\end{tabular}   &  \begin{tabular}{c}
$2\eta-\alpha$ \\ 

$2\eta$ \\ 
 
$-1$ \\ 
\end{tabular}\\

\\
\\
QDIH-$IA$ :  & \begin{tabular}{c}
$\begin{bmatrix}
 \alpha -2 \alpha  \eta ^2 -1 & -\alpha  \eta  (1-2 \eta ^2)^{\frac{1}{2}} & \alpha  \eta  (1-2 \eta ^2)^{\frac{1}{2}} \\
-\alpha  \eta  (1-2 \eta ^2)^{\frac{1}{2}} & \eta+\alpha  \eta ^2-\frac{1}{2} & \frac{1}{2}+\eta-\alpha  \eta ^2 \\
\alpha  \eta  (1-2 \eta ^2)^{\frac{1}{2}} & \frac{1}{2}+\eta-\alpha  \eta ^2 &\eta+\alpha  \eta ^2 -\frac{1}{2}
\end{bmatrix}$ \\ 
 \\
\begin{footnotesize}
$ =2\eta I_0-(1-\frac{\alpha}{2})I_1 + 2\alpha (\eta^2-\frac{1}{4})I_2 +\alpha \eta (1-2\eta^2)^{1/2} I_{3}.$
\end{footnotesize} \\ 
\end{tabular}   &  \begin{tabular}{c}
$\alpha-1$ \\ 

$-1$ \\ 
 
$2\eta$ \\ 
\end{tabular}\\
\\
\\
QDIH-$IB$ :  & \begin{tabular}{c}
$\left[
\begin{array}{ccc}
 1-\alpha +2 \alpha  \eta ^2 & \alpha  \eta  (1-2 \eta ^2)^{\frac{1}{2}} & -\alpha  \eta  (1-2 \eta ^2)^{\frac{1}{2}} \\
 \alpha  \eta  (1-2 \eta ^2)^{\frac{1}{2}} & \frac{1}{2}+\eta-\alpha  \eta ^2 & \eta+\alpha  \eta ^2 -\frac{1}{2}\\
-\alpha  \eta  (1-2 \eta ^2)^{\frac{1}{2}}  & \eta+\alpha  \eta ^2 -\frac{1}{2} & \frac{1}{2}+\eta -\alpha  \eta ^2
\end{array}
\right]$ \\ 
 \\
\begin{footnotesize}
$ =2\eta I_0+(1-\frac{\alpha}{2})I_1 - 2\alpha (\eta^2-\frac{1}{4})I_2 -\alpha \eta (1-2\eta^2)^{1/2} I_{3}.$
\end{footnotesize} \\ 
\end{tabular}   &  \begin{tabular}{c}
$1-\alpha$ \\ 

$1$ \\ 
 
$2\eta$ \\ 
\end{tabular}\\
\\
QDIH-$IC$ :  & \begin{tabular}{c}
$\left[
\begin{array}{ccc}
 1-\alpha +2 \alpha  \eta ^2 & \alpha  \eta  (1-2 \eta ^2)^{\frac{1}{2}} & -\alpha  \eta (1-2 \eta ^2)^{\frac{1}{2}}\\
\alpha  \eta  (1-2 \eta ^2)^{\frac{1}{2}}  & \frac{1}{2}-\eta-\alpha  \eta ^2 & \alpha  \eta ^2-\eta-\frac{1}{2} \\
-\alpha  \eta (1-2 \eta ^2)^{\frac{1}{2}} & \alpha  \eta ^2-\eta-\frac{1}{2} & \frac{1}{2}-\eta-\alpha  \eta ^2
\end{array}
\right]$ \\ 
 \\
\begin{footnotesize}
$ = -2\eta I_0+(1-\frac{\alpha}{2})I_1 -2\alpha (\eta^2-\frac{1}{4})I_2 -\alpha \eta (1-2\eta^2)^{1/2} I_{3}.$
\end{footnotesize} \\ 
\end{tabular}   &  \begin{tabular}{c}
$1-\alpha$ \\ 

$1$ \\ 
 
$-2\eta$ \\ 
\end{tabular}\\
\\
\hline 
\end{tabular}
\caption{\footnotesize The parametrization of $M_{\mu\tau}$ for six different QDN cases with two free parameters $(\alpha,\eta)$ with four basic building blocks $I_{i=0,1,2,3}$. $m_0$ is the input parameter.}
\label{chap2table4}
\end{table}

\section{TBM, deviation from TBM and BM mixing}
We experience that $M_{\mu\tau}$ parametrized with $(\alpha,\eta)$ (see Table.(\ref{chap2table4})) gives certain correlation between absolute masses and $\theta_{12}$  ,
\begin{align}
\label{eqq}
\sin\theta_{12}= \frac{1}{\sqrt{2}}\frac{m_2}{m_3}\,\,\text{(QDNH case)},\\
\label{eqqq}
\sin\theta_{12}= \frac{1}{\sqrt{2}}\frac{m_3}{m_2}\,\,\text{(QDIH case)}. 
\end{align}
Considering QDNH case as an example, we find,
\begin{eqnarray}
\Delta m_{21}^2 &=& \alpha(2\sqrt{2}\sin\theta_{12}-\alpha)\,m_{0}^2,\\
\Delta m_{31}^2 &=& (1-\alpha+\sqrt{2} \sin\theta_{12})(1+\alpha -\sqrt{2}\sin\theta_{12})\, m_{0}^2.
\end{eqnarray}
For all the QDNH cases, we fix the input, $m_0 = 0.082\, eV$. TBM condition implies $\theta_{12}=\sin^{-1}(1/\sqrt{3})$. A choice of the free parameter, $\alpha = 1.626$ (QDNH-$IA$), We obtain $\Delta m_{21}^2 \sim 7.6 \times 10^{-5}\, eV^2$ and $\Delta m_{31}^2 \sim 2.32 \times 10^{-3} eV^2$ \cite{Tortola:2012te,Fogli:2012ua,GonzalezGarcia:2012sz}. 

If we expect a little deviation from TBM mixing, say $\sin^2\theta_{12}=0.32$ then along with a choice of $\alpha= 1.5929$, we obtain $\Delta m_{21}^2 \sim 7.6 \times 10^{-5}\, eV^2$ and $\Delta m_{31}^2 \sim 2.49 \times 10^{-3} eV^2$\cite{Tortola:2012te}. Similar treatment holds good for the remaining cases also. The parametrization of the mass matrix with two free parameters $(\alpha,\eta)$ is compatible with both TBM mixing and with deviation from TBM as well, and agrees to the global data \cite{Tortola:2012te,Fogli:2012ua,GonzalezGarcia:2012sz}.   

But the BM mixing ($\sin\theta_{12}=1/\sqrt{2}$) is somehow disfavoured by all the six QDN mass models $M_{\mu\tau}$ ($\alpha, \eta $) (see Table.(\ref{chap2table4})). The BM mixing will lead to, $m_2=m_3$, which implies $\Delta m_{21}^2=\Delta m_{31}^2.$ (See Eqs.\,(\ref{eqq})-(\ref{eqqq}) 
Needless to mention that this is problem never arises if we adopt the general parametrization with three free parameters $(\alpha,\beta,\eta)$ (see Table.(\ref{chap2table2})).
\section{Charged lepton correction}

We derive the diagonalizing matrix for both $M_{\mu\tau}(\alpha,\beta,\eta)$ (see Table. (\ref{chap2table2})) and $M_{\mu\tau}(\alpha,\eta)$ (see Table.(\ref{chap2table4})) in the exact form as shown below,
\begin{align}
U_{\nu L}=\begin{pmatrix}
(1-2\eta^2)^{1/2} & \sqrt{2}\eta & 0 \\ 
-\eta & \frac{1}{\sqrt{2}}(1-2\eta^2)^{1/2} & \frac{1}{\sqrt{2}} \\ 
\eta  & -\frac{1}{\sqrt{2}}(1-2\eta^2)^{1/2}& \frac{1}{\sqrt{2}}
\end{pmatrix}.
\end{align}

Indeed, $\theta_{13}$ is zero and $\theta_{23}$ is $\pi/4$. We have to include some extra ingredient in order to deviate $\theta_{13}$ and $\theta_{23}$ from what $M_{\mu\tau}$ says. 

The mixing matrix in the lepton sector, $U_{PMNS}$, appears in the electro-weak coupling to the $W$ bosons and is expressed in terms of lepton mass eigenstates. We have, 
\begin{eqnarray}
\mathcal{L}=-\bar{e_L}M_e e_R -\frac{1}{2}\bar{\nu_L}m_{LL}^{\nu}\nu_{L}^c +H.c,
\end{eqnarray}
A transformation from flavour to mass basis: $U_{eL}^{\dagger}M_e U_{eR}=diag (m_e, m_\mu, m_\tau)$ and $U_{\nu L}^{\dagger}m_{LL}^{\nu} U_{\nu R}=diag(m_1,m_2,m_3)$ gives \cite{King:2002nf,Frampton:2004ud, Altarelli:2004jb, Antusch:2004re, Feruglio:2004gu, Mohapatra:2005yu,Antusch:2005kw},
\begin{eqnarray}
\label{chap2upmns}
U_{PMNS}=U_{eL}^{\dagger}U_{\nu L}.
\end{eqnarray}
As stated earlier, it was assumed $U_{eL}=I$ and hence $U_{PMNS}=U_{\nu L}(\eta)$. Probably a suitable texture of $U_{eL}$ other than $I$, satisfying the unitary condition, may give rise to the desired deviation in the mixing angles. The mixing angle, $\theta_{12}$ is controlled efficiently with $\mu-\tau$ symmetry. We want to preserve this important property even though contribution from charged lepton sector is considered.

\subsection{The charged lepton mixing matrix} 
 
In the absence of any CP phases, the charged lepton mixing matrix takes the form of a general $3 \times 3$ orthogonal matrix. In order to parametrize a $3 \times 3$ orthogonal matrix we require three rotational matrices of the following form.
\begin{eqnarray}
R_{12}(\theta)&=&\begin{bmatrix}
c_{\theta} & s_{\theta} & 0 \\ 
-s_{\theta} & c_{\theta} & 0 \\ 
0 & 0 & 1
\end{bmatrix},\\
 R_{23}(\sigma)&=&\begin{bmatrix}
1 & 0 & 0\\ 
0 & c_{\sigma} & s_{\sigma} \\ 
0 & -s_{\sigma} & c_{\sigma}
\end{bmatrix},\\
R_{23}(\sigma)&=&\begin{bmatrix}
c_{\tau} & 0 & s_{\tau}\\ 
0 &1 & 0 \\ 
-s_{\tau} & 0 & c_{\tau}
\end{bmatrix},
\end{eqnarray}
where, $s_{\omega}=\sin \omega$ and $c_{\omega}=\cos \omega$. We experience nine independent choices of combining these independent rotational matrices in order to generate the general orthogonal matrix \cite{Fritzsch:1997st}. Out of all these choices, we prefer $R=R_{12}(\theta)R_{31}(\tau)R_{23}(\sigma)$ in the charged lepton sector, which is different from the standard parametrization scheme. Again keeping in mind the fact that $R_{ij}^{-1}(\omega)$ plays an equivalent role as $R_{ij} (\omega)$ \cite{Fritzsch:1997st} in the construction of the general orthogonal matrix, we parametrize the charged lepton mixing matrix, $U_{PMNS}$ (see Eq.(\ref{chap2upmns})),
\begin{eqnarray}
\tilde{U}_{eL}=\tilde{R}_{12}^{-1}(\theta)\tilde{R}_{31}^{-1}(\sigma) \tilde{R}_{23}(\tau),
\end{eqnarray}
and along with the small angle approximation: $s_{\omega}=\omega$ and $c_{\omega}= 1-\omega^2/2$, we finally construct the PMNS matrix, of which the three important elements are,
\begin{eqnarray}
&U_{e2}  \approx  s_{12}^{\nu} + \frac{c^{\nu}_{12}}{\sqrt{2}}(\theta-\sigma)-\frac{s_{12}^{\nu}}{2}(\theta^2 +\sigma^2),&\\
&U_{e3}  \approx   \frac{1}{\sqrt{2}} (\theta +\sigma),&\\
&U_{\mu 3}  \approx    \frac{1}{\sqrt{2}}-\frac{1}{\sqrt{2}}\tau-\frac{1}{2\sqrt{2}}(\theta^2+\sigma^2),&
\end{eqnarray}

\begin{figure}
\begin{center}
\includegraphics[scale=0.74]{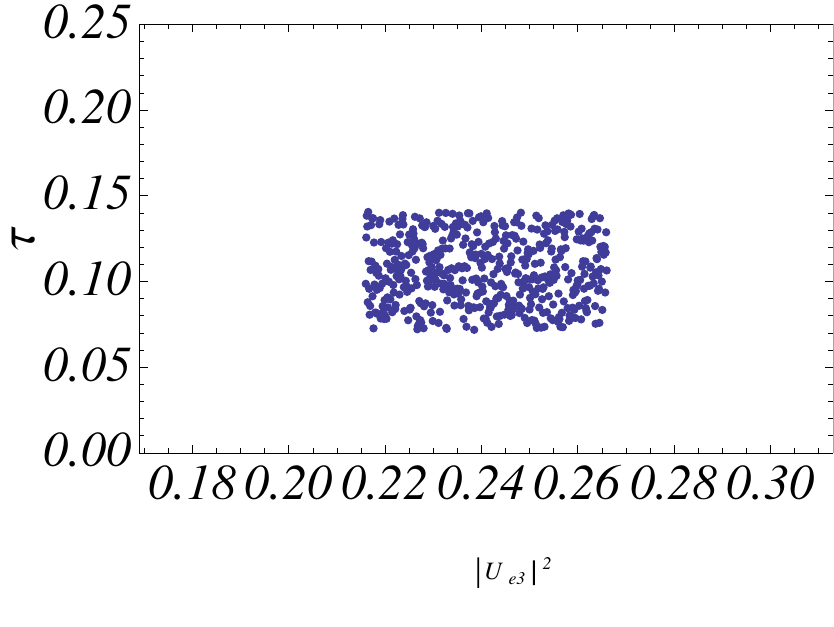}
\caption{\footnotesize Graphical analysis to fix the parameter, $\tau$ against the $1\sigma$ range of $\sin^2\theta_{13}=|U_{e3}|^2$. } 
\label{chap2fig4}
\end{center}
\end{figure}

where $s_{12}^{\nu}=\sqrt{2}\eta$ and $c_{12}^{\nu}=\sqrt{1-2\eta^2}$. The choice of the $\sigma$, $\theta$ and $\tau$ are arbitrary. So that $\sin\theta_{12}$ as obtained from $M_{\mu\tau}$ is not disturbed, the middle term in the expression of $U_{e2}$ must vanish, $\theta-\sigma=0$. We choose, $\theta,\,\sigma=\lambda/2$, ($\lambda=0.2253\pm0.0007$, standard Wolfenstein parameter\cite{Wolfenstein:1983yz}) and get $\sin\theta_{13}=|U_{e3}|=\lambda/\sqrt{2}$ \cite{Antusch:2012fb}. Once, $\theta$ and $\sigma$ are fixed, the choice of $\tau$ is guided by the requirement of necessary deviation of $\theta_{23}$ from the maximal condition. We see, in Fig (\ref{chap2fig4}). with respect to $1\sigma$ range of $|U_{e3}|^2$, $\tau$ centers around $\tau\sim 0.1 \sim \lambda/2$. Finally we model 
\begin{eqnarray}
 \tilde{U}_{eL}=\tilde{R}_{12}^{-1}(\lambda/2)\tilde{R}_{31}^{-1}(\lambda/2) \tilde{R}_{23}(\lambda/2). 
 \end{eqnarray}
so that,
\begin{align}
\tilde{U}_{eL}^{\dagger} &\approx \left[
\begin{array}{ccc}
 1-\frac{\lambda ^2}{4} & \frac{\lambda }{2} & \frac{\lambda}{2}\\
 -\frac{\lambda}{2} +\frac{\lambda ^2}{4} & 1-\frac{\lambda ^2}{4} & -\frac{\lambda}{2} \\
 -\frac{\lambda}{2}-\frac{\lambda ^2}{4}  & \frac{\lambda}{2}-\frac{\lambda ^2}{4} & 1-\frac{\lambda ^2}{4}
\end{array} \right]+\mathcal{O}(\lambda^3).
\end{align} 
 
\subsection{Breaking the $\mu-\tau$ interchange symmetry}

Once, the charged lepton contributions are taken into consideration the $\mu-\tau$ symmetry will be perturbed. Finally, we obtain, 
the corrected neutrino mass matrix, $m_{LL}^{\nu}(\alpha,\eta,\lambda) = \tilde{U}_{eL}^{\dagger}. M_{\mu\tau} \tilde{U}_{eL}$. The invariant building blocks $I_{i=0,1,2,3}$ (see Table.(\ref{chap2building})) of $M_{\mu\tau}$ will now change to,
\begin{align}
I^\lambda_{i=0,..,3}&= \tilde{U}_{eL}^{\dagger}.I_{i=0,..,3}. \tilde{U}_{eL}\nonumber\\
&=I_{i=0,..,3}+\Delta I_{i=0,..,3}^{\lambda}+\mathcal{O}(\lambda^3). 
\end{align} 
The  matrices $\Delta I_{i}^{\lambda}$ s are listed in Table.(\ref{chap2table5}).
\begin{table}
\setlength{\tabcolsep}{0.06em}
\begin{small}
\centering
\begin{tabular}{ll}
\hline
\hline
 & $\Delta I_{i}^{\lambda}$ \\ 
\hline 
\hline
\\
$\Delta I_{0}^{\lambda}\quad\approx$ & $\frac{1}{2}\lambda \left[
\begin{array}{ccc}
 \lambda & 1-\frac{1}{2}\lambda & 1+\frac{1}{2}\lambda \\
 1-\frac{1}{2}\lambda & -1-\frac{1}{4}\lambda & -\lambda \\
 1+\frac{1}{2}\lambda & -\lambda & 1-\frac{3}{4}\lambda
\end{array}
\right]$ \\ 

$\Delta I_{1}^{\lambda}\quad=$ & $\quad-\Delta I_{0}^{\lambda}$  \\ 

$\Delta I_{2}^{\lambda}\quad\approx$ & $\frac{1 }{2}\lambda \left[
\begin{array}{ccc}
\lambda  & -\frac{1}{2}\lambda & 1+\frac{1}{2}\lambda \\
 -\frac{1}{2}\lambda  & 1-\frac{3}{4} \lambda & 0 \\
-\frac{1}{2}\lambda & 0 & -1-\frac{1}{4}\lambda
\end{array}
\right] $ \\ 

$\Delta I_{3}^{\lambda}\quad\approx$ & $ \frac{1}{2}\lambda \left[
\begin{array}{ccc}
 0 & -1+\lambda & -1-\frac{1}{2}\lambda\\
 -1+\lambda & 2 & 2\lambda  \\
 -1-\frac{1}{2}\lambda& 2\lambda & -2
\end{array}
\right]$ 
\\
\\ 
\hline
\hline

& $m_{LL}^{\nu}(\alpha,\eta,\lambda)$\\
\hline
\hline
\\

QDNH-$IA$ : & \begin{scriptsize}
$(I_0+\Delta I_{0}^{\lambda})-(2\eta- \frac{\alpha}{2})(I_1-\Delta I_{0}^{\lambda}) +  2\alpha (\eta^2-\frac{1}{4})(I_2+\Delta I_{2}^{\lambda})
 +\alpha \eta (1-2\eta^2)^{1/2} (I_{3}+\Delta I_{3}^{\lambda})$
\end{scriptsize}\\
\\
QDNH-$IB$ : & \begin{scriptsize}
$ (I_0+\Delta I_{0}^{\lambda})+(2\eta-\frac{\alpha}{2})(I_1-\Delta I_{0}^{\lambda}) -2\alpha (\eta^2-\frac{1}{4})(I_2+\Delta I_{2}^{\lambda}) -\alpha \eta (1-2\eta^2)^{1/2} (I_{3}+\Delta I_{3}^{\lambda})$
\end{scriptsize} \\
\\
QDNH-$IC$ : &\begin{scriptsize}
$-(I_0+\Delta I_{0}^{\lambda})+(2\eta-\frac{\alpha}{2})(I_1-\Delta I_{0}^{\lambda}) -2\alpha (\eta^2-\frac{1}{4})(I_2\Delta + I_{2}^{\lambda}) -\alpha \eta (1-2\eta^2)^{1/2} (I_{3}+\Delta I_{3}^{\lambda})$
\end{scriptsize} \\
\\
QDIH-$IA$ : & \begin{scriptsize}
$ 2\eta (I_0+\Delta I_{0}^{\lambda})-(1-\frac{\alpha}{2})(I_1-\Delta I_{0}^{\lambda}) + 2\alpha (\eta^2-\frac{1}{4})(I_2+\Delta I_{2}^{\lambda}) +\alpha \eta (1-2\eta^2)^{1/2} (I_{3}+\Delta I_{3}^{\lambda})$
\end{scriptsize}\\
\\
QDIH-$IB$ : &\begin{scriptsize}
$ 2\eta (I_0+\Delta I_{0}^{\lambda})+(1-\frac{\alpha}{2})(I_1-\Delta I_{0}^{\lambda}) - 2\alpha (\eta^2-\frac{1}{4})(I_2+\Delta I_{2}^{\lambda}) -\alpha \eta (1-2\eta^2)^{1/2} (I_{3}+\Delta I_{3}^{\lambda}).$
\end{scriptsize} \\
\\
QDIH-$IC$ : &\begin{scriptsize}
$  -2\eta (I_0+\Delta I_0^{\lambda})+(1-\frac{\alpha}{2})(I_1-\Delta I_{0}^{\lambda}) -2\alpha (\eta^2-\frac{1}{4})(I_2+\Delta I_2^{\lambda}) -\alpha \eta (1-2\eta^2)^{1/2} (I_{3}+\Delta I_3^{\lambda}).$
\end{scriptsize}
\\ 
\\
\hline
\hline
\\
$U_{PMNS}\approx$ & \begin{tabular}{l}

$\begin{footnotesize} 
\begin{bmatrix}
c_{12}^{\nu }(1-\frac{1}{4} \lambda ^2)&s_{12}^{\nu }(1-\frac{1}{4} \lambda ^2) & \frac{\lambda}{\sqrt{2}}\\
-\frac{s_{12}^{\nu}}{\sqrt{2}}-\frac{\lambda}{2}(1-\frac{\lambda}{2})(\frac{s_{12}^{\nu}}{\sqrt{2}}+c_{12}^{\nu}) &\frac{c_{12}^{\nu}}{\sqrt{2}}+\frac{\lambda}{2}(1-\frac{\lambda}{2})(\frac{c_{12}^{\nu}}{\sqrt{2}}-s_{12}^{\nu})&\frac{1}{\sqrt{2}}-\frac{\lambda}{2\sqrt{2}}\\
\frac{1}{\sqrt{2}}(1-\frac{\lambda}{2})s_{12}^{\nu}-\frac{\lambda}{2}(1+\frac{\lambda}{2})c_{12}^{\nu}& -\frac{1}{\sqrt{2}}(1-\frac{\lambda}{2})c_{12}^{\nu}-\frac{\lambda}{2}(1+\frac{\lambda}{2})s_{12}^{\nu}&\frac{1}{\sqrt{2}}+\frac{\lambda}{2\sqrt{2}}
\end{bmatrix},\end{footnotesize} $\\ 
\end{tabular} \\
\\
&${\footnotesize s_{12}^{\nu}=\sqrt{2}\eta,\,\,c_{12}^{\nu}=(1-2\eta^2)^{1/2}}$ \\
\\
\hline
\end{tabular} 
\caption{\footnotesize The perturbation to the respective building block matrices, $I_{i}$s are estimated in terms of $\Delta I_{i}$s. The corresponding textures of the corrected mass matrices $m_{LL}^{\nu}(\alpha,\eta,\lambda)$ are also described. The lepton mixing matrix which is now modified from $U_{\nu L}$ to $U_{eL}^{\dagger}.U_{\nu L}$ is also presented. } 
\label{chap2table5}
\end{small}
\end{table}
We consider the case of $M_{\mu\tau}$ with two free parameters $(\alpha,\eta)$ for QDNH-$TypeA$ case (see Eq.(\ref{chap2QNHAA})), as example.
\begin{eqnarray}
 m_{LL}^{\nu}(\alpha,\eta,\lambda)&=&\tilde{U}_{eL}^{\dagger}. M_{\mu\tau}(\alpha,\eta) \tilde{U}_{eL} \nonumber\\
&=& I_0^{\lambda}-(2\eta- \frac{\alpha}{2})I^{\lambda}_1 +  2\alpha (\eta^2-\frac{1}{4})I_2^{\lambda}+\alpha \eta (1-2\eta^2)^{1/2} I_{3}^{\lambda},\nonumber\\
&=&
\begin{bmatrix}
\alpha-2\eta-2 \alpha \eta^2 & -\alpha \eta (1-2 \eta^2)^{\frac{1}{2}} & \alpha \eta (1-2 \eta^2)^{\frac{1}{2}} \\ 
-\alpha \eta (1-2 \eta^2)^{\frac{1}{2}} & \frac{1}{2}-\eta+\alpha \eta^2 & \frac{1}{2} +\eta -\alpha \eta^2 &\\ 
\alpha \eta (1-2 \eta^2)^{\frac{1}{2}}  & \frac{1}{2} +\eta -\alpha \eta^2 & \frac{1}{2}-\eta+\alpha \eta^2 &
\end{bmatrix}\nonumber\\
&&\quad+
\frac{\lambda}{2}\left(1-2\eta-\frac{\alpha}{2}\right) \left[
\begin{array}{ccc}
 \lambda & 1-\frac{1}{2}\lambda & 1+\frac{1}{2}\lambda \\
1-\frac{1}{2}\lambda & -1-\frac{1}{4}\lambda & -\lambda \\
1+\frac{1}{2}\lambda  & -\lambda & 1-\frac{3}{4}\lambda
\end{array}
\right]\nonumber\\
&&\quad+ 
\alpha\lambda \left(\eta^2-\frac{1}{4}\right) \left[
\begin{array}{ccc}
\lambda  & -\frac{\lambda}{2} & 1+\frac{1}{2}\lambda \\
 -\frac{\lambda}{2}  & 1-\frac{3}{4} \lambda & 0 \\
 1+\frac{1}{2}\lambda & 0 & -1-\frac{1}{4}\lambda
\end{array}
\right]
\nonumber\\
&&\quad+
\frac{\alpha \eta\lambda}{2} (1-2\eta^2)^{1/2} \left[
\begin{array}{ccc}
 0 & -1+\lambda & -1-\frac{1}{2}\lambda\\
 -1+\lambda & 2 & 2\lambda  \\
-1-\frac{1}{2}\lambda & 2\lambda & -2
\end{array}
\right]+\mathcal{O}(\lambda^3)\nonumber\\
\end{eqnarray}
The details of the texture for other QDN cases are described in Table.(\ref{chap2table5}). The texture of the PMNS matrix, $U_{PMNS}=U_{eL}^{\dagger}.U_{\nu L}$, is presented in Table.(\ref{chap2table5}). We obtain,
\begin{eqnarray}
\sin^2\theta_{12}&=& 2\eta^2+\mathcal{O}(\lambda^3),\\
\sin^2\theta_{13}&=&\frac{1}{2}\lambda^2+\mathcal{O}(\lambda^3),\\
\sin^2\theta_{23}&=&\frac{1}{2}-\frac{1}{2}\lambda+\frac{1}{8}\lambda^2+\mathcal{O}(\lambda^3).
\end{eqnarray}

\begin{table}
\begin{footnotesize}
\setlength{\tabcolsep}{0.7 em}
\begin{tabular}{l l l l l l l}
\hline
\hline 
QD & NH-IA & NH-IB &  NH-IC &  IH-IA &  IH-IB &  IH-IC \\ 
\hline 
\hline
\\
$\alpha$ (TBM) & 1.626 & 0.0068 & 0.0068 & 1.9946 & 0.0054 & 0.0054 \\ 
\\
$\alpha$  & 1.5929 & 0.0071 & 0.0071 & 1.9946 & 0.0054 & 0.0054 \\ 
\\ 
$\eta$ (TBM) & 0.4083 & 0.4083 & 0.4083 & 0.4083 & 0.4083 & 0.4083 \\ 
\\ 
$\eta$ & 0.40 & 0.40 & 0.40 & 0.3987 & 0.3987 & 0.3987\\ 
\\
$m_{0}$ $eV$ & 0.082 & 0.082 & 0.082 &  0.084 & 0.084 & 0.084\\ 
\\
\hline 
$m_{1}$ $eV$(TBM) & 0.06638 & 0.06639 & 0.06639 & 0.08355 & 0.08355 & 0.08355 \\ 
\\ 
$m_{2}$ $eV$(TBM) & - 0.06695 & 0.06695 & 0.06695 & -0.084 & 0.084 & 0.084 \\ 
\\ 
$m_{3}$ $eV$(TBM) & 0.082 & 0.082 & -0.082 & 0.06859 & 0.06859 & -0.06859 \\ 
\\ 
$m_{1}$ $eV$ & 0.06502 &  0.06502 & 0.06502 & 0.08355 & 0.08355 & 0.08355 \\ 
\\ 
$m_{2}$ $eV$ & -0.0656 & 0.0656 & 0.0656 & -0.084 & 0.084 & 0.084 \\ 
\\ 
$m_{3}$ $eV$ & 0.082 & 0.082 & -0.082 & 0.0672 & 0.0672 & -0.0672 \\ 
\\ 
$\Delta m_{21}^2(10^{-5}eV^2)$ (TBM) & 7.645 & 7.435  & 7.435 & 7.60 &  7.60 &  7.60 \\ 
\\ 
$\Delta m_{31}^2(10^{-3}eV^2)$ (TBM) & 2.318 & 2.316 & 2.316 & -2.352 & -2.28 & -2.28 \\ 
\\ 
$\Delta m_{21}^2(10^{-5}eV^2)$  & 7.605 & 7.605 & 7.605 & 7.60 & 7.60 & 7.60 \\ 
\\ 
$\Delta m_{21}^2(10^{-3}eV^2)$  & 2.497 & 2.497 & 2.497 & -2.464 & -2.464 & -2.464\\ 
\\ 
$\Sigma\, m_{i}$ $eV$ (TBM) & 0.2153 & 0.2154& 0.2154 &  0.23613 & 0.23613 & 0.23613\\ 
\\ 
$\Sigma\, m_{i}$ $eV$ & 0.21262 & 0.21262 & 0.21262 &  0.23475 & 0.23475 & 0.23475\\ 
\\ 
$\sin^2\theta_{12}$ & 0.319 & 0.319 &  0.319 & 0.3195 & 0.3195 & 0.3195\\ 
\\ 
$\sin^2\theta_{13}$& 0.0252 & 0.0252 & 0.0252 & 0.0252 & 0.0252 & 0.0252 \\ 
\\ 
$\sin^2\theta_{23}$ & 0.3943 & 0.3943 & 0.3943 & 0.3943 & 0.3943 & 0.3943 \\ 
$m_{\nu_{e}}\, eV$ & 0.06582       &  0.06582      &   0.06582   &  0.0835      & 0.0835       &  0.0835                 \\
$m_{ee}\, eV     $ & 0.02452     &  0.06590        &   0.06174   &   0.03063     & 0.083625       &     0.08021       \\
\\
\hline
\end{tabular} 
\caption{\footnotesize The study of the six cases of Quasi degenerate neutrino mass model for both TBM mixing and deviation from TBM mixing. The analysis is done with the parameters ($\alpha,\, \eta, \lambda $) and input $m_0$. $m_0$ is fixed at $0.082\, eV$ (QDNH) and $0.084\, eV$ (QDIH) respectively. The free parameter $\alpha$ is related with absolute masses. The free parameter $\eta$ controls both masses and the solar angle. $\lambda=0.2253$, the Wolfenstein parameter is related with deviation of reactor angle from zero and that for atmospheric from maximal condition.}
\label{chap2table6}
\end{footnotesize}
\end{table}

\section{Numerical calculation}

We assign certain ranges to the free parameter $\alpha$ and $\eta$ respectively. Based on the $1\sigma$ range of the physical observable quantities available from Global data analysis \cite{Fogli:2012ua}, we assign $\alpha = 1.5939-1.6239$ \text{(QDNH-IA)},$\,0.0080-0.0220$ (QDNH-IB,IC), $1.9945-1.9948$ (QDIH-IA), $0.0052-0.0055$ (QDIH-IB,IC) , and $\eta= 0.3814-0.4031$. The input parameter  $m_0 \sim 0.08 \, eV.$ and  $\lambda=0.2253$. We have now four parameters, out of which $\alpha$ and $\eta$ are free and the number of unknowns present is six.

\subsection{Observable parameters in oscillation experiments and cosmological observation}

We apply the six QDN neutrino mass matrices $m_{LL}^{\nu}(\alpha,\eta,\lambda)$ to study their relevance in the oscillation experiments. It is found that under a suitable choice of the free parameters ($\alpha,\eta$), all the six QDN models are equally capable of describing both TBM and TBM-deviated scenarios (see Table.(\ref{chap2table6})) and are indistinguishable. QDNH model says, $|m_1|, |m_2|\sim 0.06\,eV$, $|m_{3}|\sim 0.08\, eV$, while $|m_2|, |m_3|\sim 0.08\,eV$, $m_{1}\sim 0.06\, eV$ for QDIH case. For both the cases, $\Delta m_{21}^2\sim 7.6\times 10^{-5}\,eV^2$ and $\Delta m_{31}^2\sim 2.4\times 10^{-3}\,eV^2$. The mixing angle parameters are $\sin^{2}\theta_{13}\simeq 0.025$, $\sin^{2}\theta_{12}\simeq 0.32$ and $\sin^{2}\theta_{23}\simeq 0.39$. Also $\Sigma |m_i|\simeq 0.21\, eV$ (QDNH case) and $\Sigma |m_i|\simeq 0.23\, eV$ (QDIH case).

We  study a quantity $\sqrt{\Delta m_{21}^2} / \sqrt{\Delta m_{31}^2}$, which according to the global data analysis lies near to $0.2$. The correlation plots in the plane $\sqrt{\Delta m_{21}^2} / \sqrt{\Delta m_{31}^2}$ and $\sin^2\theta_{12}$ for all QDN models are shown in Fig (\ref{chap2fig4}). We see for QDNH-Type IA case, there exists a sharp bound on $\sin^2\theta_{12}$ around $0.32$ which is the experimental best-fit of $\sin^2\theta_{12}$ according to Global data analysis \cite{Tortola:2012te}.  
\begin{figure}
\begin{center}
\includegraphics[scale=0.55]{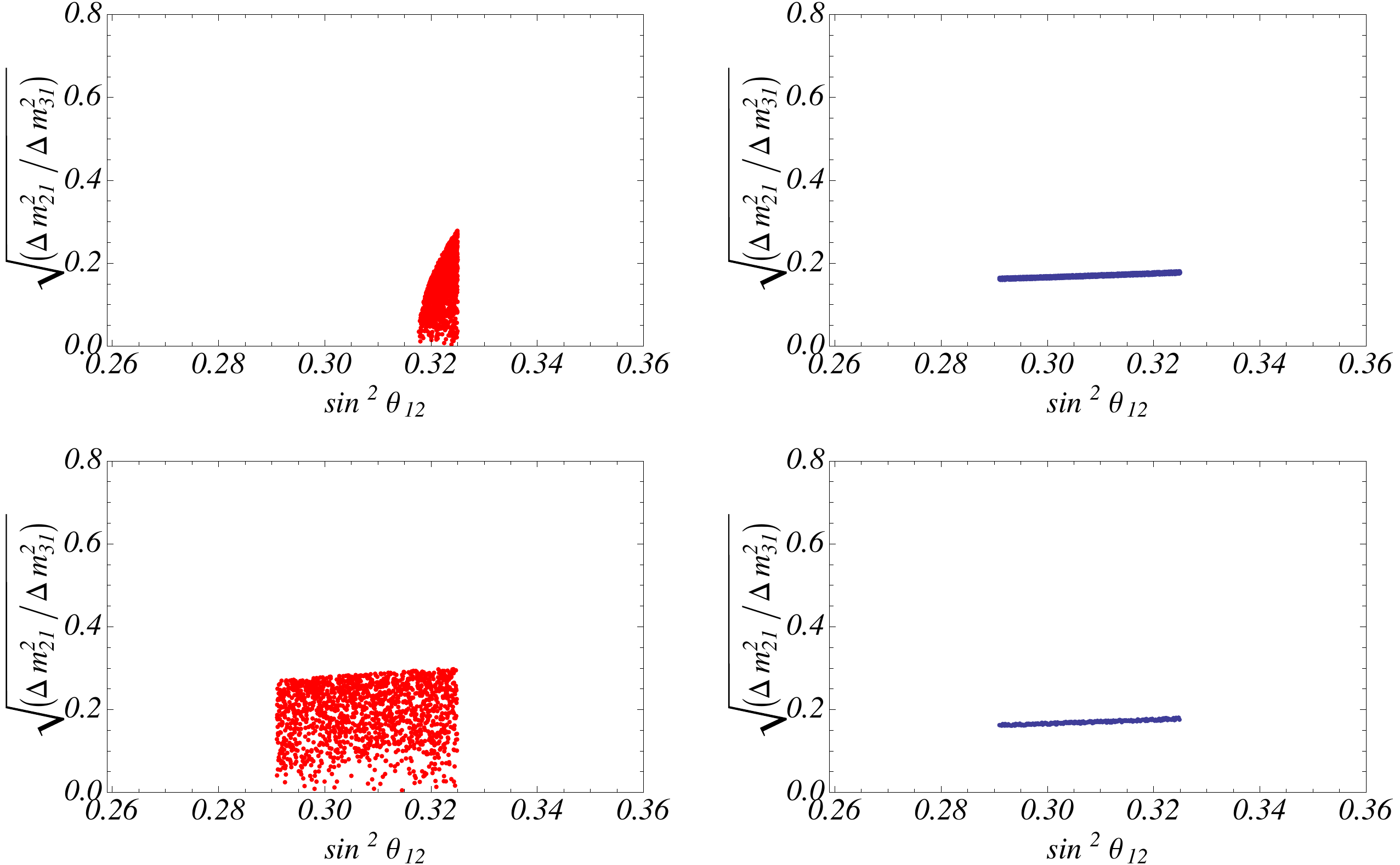} 
\caption{\footnotesize The correlation plots in the plane of $\sqrt{\Delta m_{21}^{2}/\Delta m_{31}^{2}}$ and $\sin^2\theta_{sol}$ for different cases of QDNH-IA (top-left), QDNH-IB,IC (bottom-left), QDIH-IA (top-right) and QDIH-IB,IC (bottom-right). The bounds on $\sqrt{\Delta m_{21}^{2}/\Delta m_{31}^{2}}$ are found to be sharp for QDIH cases. The experimental value of this quantity must lie close to 0.2. For QDNH-IA case, we obtain a bound on $\sin^2\theta_{sol}$ around a value of $0.32$.  } 
\label{chap2fig5}
\end{center}
\end{figure}

\subsection{Absolute electron neutrino mass ($m_{\nu_{e}}$) and Effective Majorana neutrino mass ($m_{ee}$)}

Besides the oscillation experiments and the cosmological bound on $\Sigma |m_{i}|$, There are other two important quantities :  effective electron neutrino mass, $m_{\nu_{e}}$ appearing in $\beta$-decay and effective Majorana mass $m_{ee}$, appearing in neutrino-less double $\beta$-decay experiment and are useful for the study of nature of the neutrino masses. 
\begin{eqnarray}
&m_{\nu _{e}} = (\Sigma m_{i}^2 |U_{ei}|^2)^{1/2},&\\
&m_{ee}= | m_{1} |U_{e1}|^2 + m_{2} |U_{e2}|^2 +  m_{3} |U_{e3}|^2|.&
\end{eqnarray}     
The results of Mainz \cite{Kraus:2004zw} and Toitsk\cite{Lobashev:1999tp} Tritium $\beta$-decay experiments,we obtain, $m_{\nu _{e}} < 2.2 \, eV$. The upcoming KATRIN experiment \cite{Weinheimer:2007zz}, expects the sensitivity upto $m_{\nu _{e}} \sim 0.3 \, eV$. In the present work, the QDNH and QDIH models predict, $m_{\nu _{e}} \sim 0.07\, eV$ and $m_{\nu _{e}} \sim 0.08\, eV$ respectively.

\begin{figure}
\begin{center}
\includegraphics[scale=0.68]{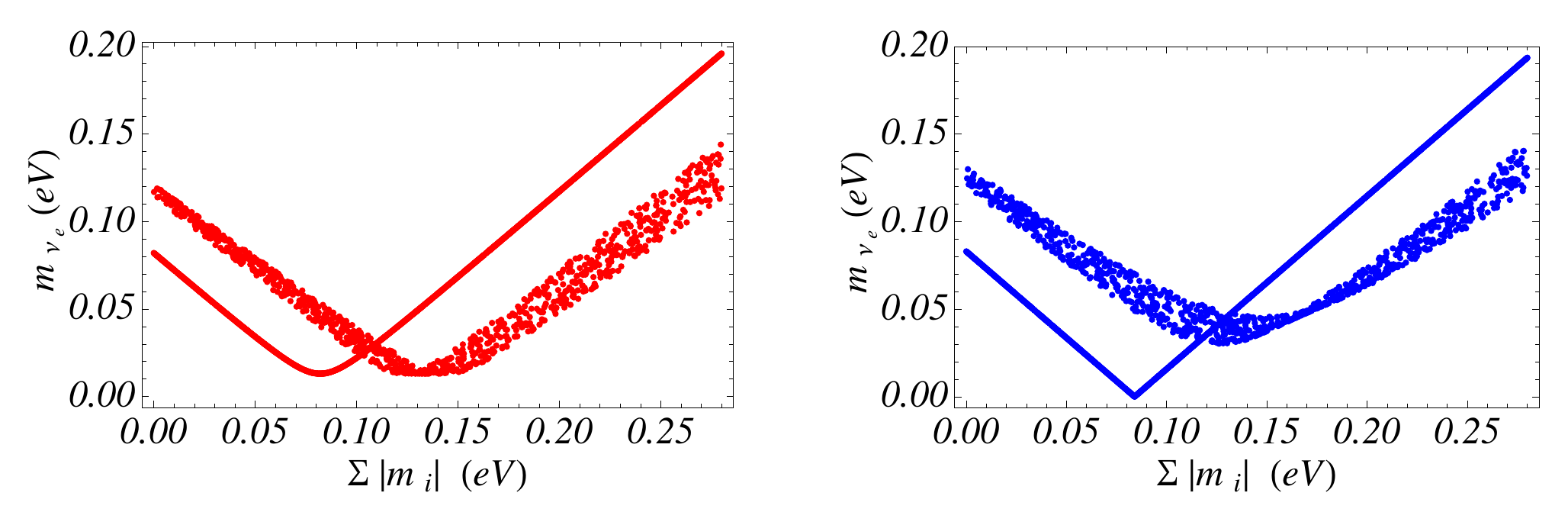}
\caption{\footnotesize A study of the correlation in the plane of $m_{\nu_{e}}$ and $\Sigma m_{i}$. Left: QDNH case, Right: QDIH case.} 
\label{chap2fig6}
\end{center}
\end{figure}

\begin{figure}
\begin{center}
\includegraphics[scale=0.75]{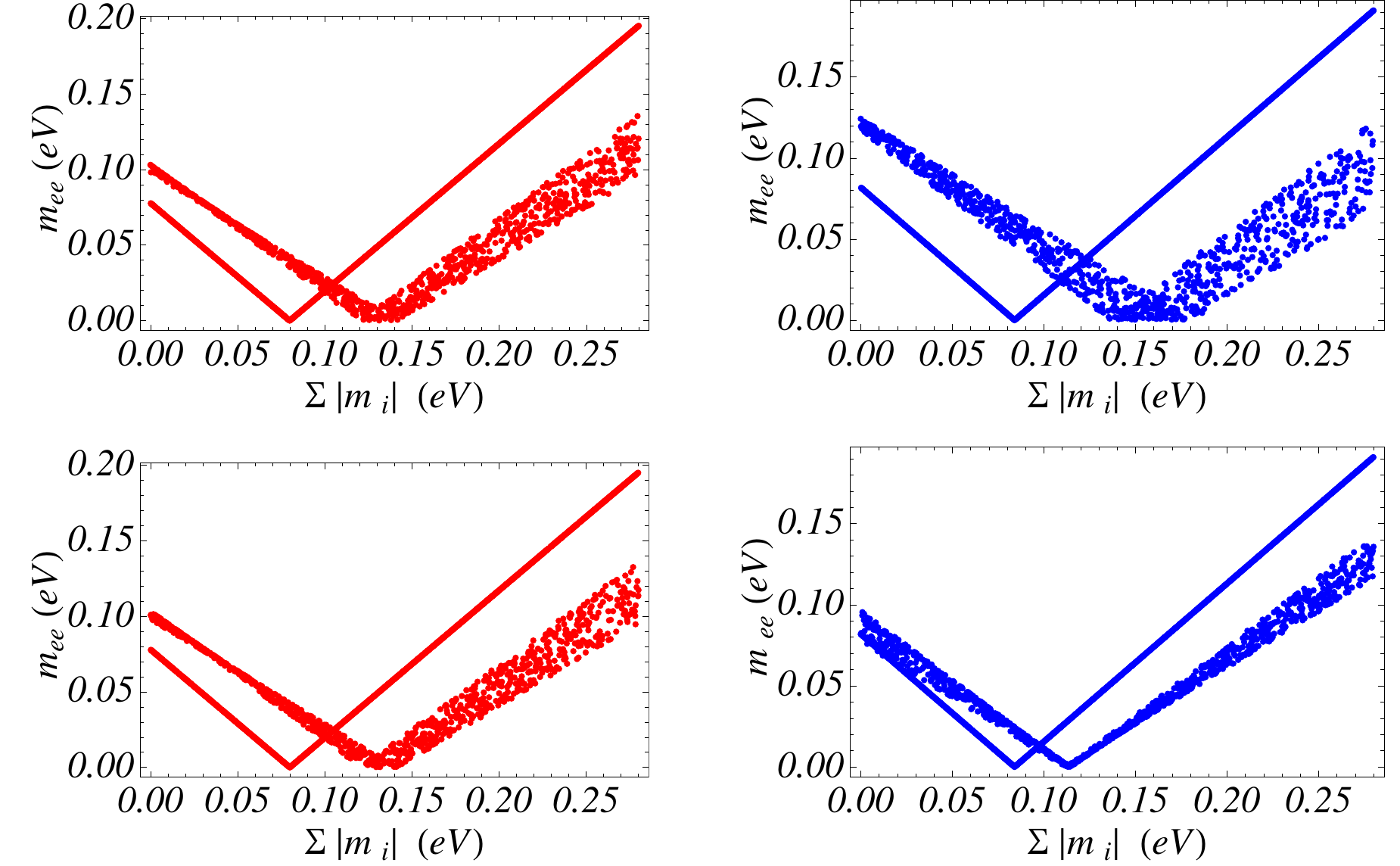}
\caption{\footnotesize  A study of the correlation in the plane of $m_{ee}$ and $\Sigma m_{i}$. Top-left: QDNH-Type IA case; top-right: QDNH-Type IB, IC cases; bottom-left: QDIH-Type IA case; bottom-right: QDIH-Type IB, IC cases. } 
\label{chap2fig7}
\end{center}
\end{figure}

The HM group \cite{KlapdorKleingrothaus:2000sn, KlapdorKleingrothaus:2004wj, KlapdorKleingrothaus:2006ff} and IGEX \cite{Aalseth:1999ji,Aalseth:2002rf,Aalseth:2002dt} groups reported the upper limit of $m_{ee}$ to $0.3-1.3\,eV$. The CUORICINO\cite{Arnaboldi:2008ds} experiment gives an improved upper bound on $m_{ee}$, $m_{ee}<0.23-0.85\,eV$. This is still considered somewhat controversial\cite{Aalseth:2002dt,Elliott:2004hr}, and requires independent confirmation. The experiments such as CUORE \cite{Arnaboldi:2002du, Fiorini:1998gj}, GERDA\cite{Schonert:2005zn}, NEMO\cite{Nasteva:2008pa,Flack:2008tf,Daraktchieva:2009mn} and Majorana\cite{Aalseth:2004yt, Avignone:2007js} will attempt to improve the sensitivity of the measurement  down to about $m_{ee}\simeq (0.05-0.09)\,eV$. Hence in that respect the QDN models are of immense importance. In our present work, QDNH and QDIH-both Type IB and Type IC models predict $m_{ee}\simeq 0.06\,eV$ and $m_{ee}\simeq 0.08\,eV$  respectively. The predictions given by Type IA cases of both QDNH and QDIH models are interesting in the sense that they leave a scope for the future experiments to go down upto a sensitivity of $m_{ee}\simeq 0.02\,eV$ and $m_{ee}\simeq 0.03\,eV$ respectively. The correlation plots are studied in the plane $m_{\nu_{e}}$ ( and $m_{ee}$ ) and $\Sigma |m_i|$ ( Fig (\ref{chap2fig6}) and Fig (\ref{chap2fig7})). 

\section{Discussion: How to discriminate different QDN models?}

We have tried to bring all the eleven cases involving six QDN, three NH and two IH cases under same roof of parametrization by introducing four common independent building block matrices, $I_{i=0,1,2,3}$. The idea of fragmentation is guided by the quest of some mechanism to save the internal texture of $M_{\mu\tau}$ against the changing solar angle. The $I_{i}$s when incorporated with the free parameters in a proper way, lead to an important feature of $M_{\mu\tau}$,  $\sin\theta_{12}=\sqrt{2}\eta$. $\theta_{12}$ is expressible in terms single parameter only, unlike the general $M_{\mu\tau}=M_{\mu\tau}(x,y,z,w)$ where $\theta_{12}$ requires the knowledge of four free parameters $(x,y,z,w)$ (see Eq.(\ref{chap2tan21})). This is also interesting to note that one of the building blocks, $I_3$ has got the same eigenstates as predicted by BM mixing (see Table.(\ref{chap2building})). The existence of this invariant textures within the mass matrix seems to be relevant and we hope that a fruitful investigation is subjected to the study of underlying discrete flavour symmetry groups.  
 
Charged lepton correction is considered as a significant tool in order to break the $\mu-\tau$ symmetry \cite{King:2002nf,Frampton:2004ud, Altarelli:2004jb, Antusch:2004re, Feruglio:2004gu, Mohapatra:2005yu,Antusch:2005kw}. The models where $\theta_{13}^{\nu}$ is very small, contributions to $\theta_{13}$ is implemented mostly from charged lepton sector. Also, this tool is very important for those models where $\theta_{23}=\pi/4$ and it may provide consistency with the LMA MSW solutions \cite{King:2013eh}. In GUT scenarios also, one finds in addition to the breaking of $ \mu-\tau$ symmetry in the neutrino sector, charged lepton corrections are unavoidable \cite{Cooper:2012wf,Hagedorn:2012ut}. Regarding the parametrization of $U_{eL}$, we have followed a parametrization scheme different from that of standard one. This step is motivated by the fact that a particular choice of parametrization does not affect the final observables, but a suitable choice can make the mathematics easier. The parametrization of $U_{eL}$ respects the GUT motivated, new QLC relation, $\theta_{13}\sim \theta_c/\sqrt{2}$ \cite{Antusch:2012fb}. In our parametrization $U_{eL}$ does not affect the prediction of $\theta_{12}$ from neutrino sector.

Our work started with the following motivations. They are, 

(1) Whether QDN neutrino mass models are equally possible like that of NH and IH models? 

(2) How to discriminate the  QDN models?

In the background of the oscillation experiments, we have tried to answer to the first question by testing the efficiency of each $m_{LL}^{\nu} (\alpha,\eta,\lambda)$ in predicting the values of five observational parameters and comparing those with Global data. In that context we find the existence of QDN neutrinos of both NH and IH pattern is undisputed. Above all, all the six QDN models sound equally possible (see Table.(\ref{chap2table6})). Hence only the oscillation experiments are not sufficient enough to answer to the second question. But here we want to mention that  QDNH-Type A model shows a strong preference for $\sin^2\theta_{12}=0.32$, which is the best-fit result according to Global analysis done by Forero $et.al$ \cite{Tortola:2012te}, evident from the correlation plot in Fig(\ref{chap2fig5}).

We have tried to find out the answer of the second question in the framework of $\beta$-decay and $0\nu\beta\beta$-decay experiments. But all the six QDN models predict the quantities $m_{\nu_{e}}$ and $m_{ee}$, below the upper bounds of the past experiments and interestingly they are much closer to the sensitivities expected to be achieved in the future experiments. In our analysis QDNH-$Type-A$ model leaves a scope for future experiments to go down upto $m_{ee}\sim 0.02 \, eV$. 

In section (\ref{chap2m0}), it has been shown that QDN nature of neutrinos permits the mass scale, $0.05\leq m_{0}\leq0.1 \,eV$. But, concerning a fair degree of degeneracy, the range is modified to, $0.07 \leq m_{0}\leq0.1 \,eV$. The ansatz regarding the correlation, $\beta: \eta \simeq 2$ plays an important role in the transition from $M_{\mu\tau}(\alpha,\beta,\eta)\rightarrow M_{\mu\tau}(\alpha,\eta) $. This ansatz holds good for the mass scale $m_0\sim 0.07-0.09\, eV$, over the $3\sigma$ range of $\beta$ and $\eta$. If there are three free parameters ($\alpha,\beta,\eta$) present in $m_{LL}^{\nu}$, degrees of arbitrariness is also quite higher. Although this ansatz restricts the arbitrariness of $\theta_{12}$ to some extent, yet only two free parameters $\alpha$ and $\eta$ (with $\lambda=0.2253$ and input $m_{0}$ fixed at $0.08\,eV$) are sufficient to predict  five observational parameters (related with oscillation experiments), in close agreement with that of experimental $1\sigma$ range of data. The parametrization respects both TBM and small TBM-deviated cases. In this context the anstaz $\beta=2\eta$, appears to be relevant and natural .

We hope that perhaps the cosmological upper bound on $\Sigma m_{i}$ have some relevance with the discrimination of the six models.   So long we adhere to $\Sigma m_{i}<0.28\,eV$\cite{Thomas:2009ae}, both QDNH and QDIH models are safe which predict, $\Sigma m_{i}=0.212$ and $\Sigma m_{i}=0.235$ respectively for input mass scale $m_0\sim 0.08 \,eV$ . But the recent analysis supports for a tighter upper bound, $\Sigma m_{i}<0.23\,eV$ \cite{Ade:2013zuv}. If so, the QDIH model seems to be insecure in our analysis. If we believe the ansatz, $\beta:\eta=2$ to be natural, then with lowering the mass scale from $0.08\,eV$ , and controlling $\alpha$ and $\eta$, we can achieve $\Sigma m_{i}<0.23\,eV$ for QDIH case, and also this will favor the TBM deviated condition. But at the same time, it will give rise to a serious problem that QDIH model with $m_0<0.08 \,eV$ will completely discard $\theta=\sin^{-1}(1/\sqrt{3})$ (TBM), because corresponding to that angle, $\Delta m_{31}^2$ will be outside the $3\sigma$ range. But the solar angle $\theta=\sin^{-1}(1/\sqrt{3})$, is still relevant within $1\sigma$\cite{Tortola:2012te}or $2\sigma$\cite{Fogli:2012ua} range. So on this basis can we discard QDIH model?  But we hope it will be too hurry to come to any conclusion. There is a possibility that by assuming $\beta:\eta=c$, where $c\neq 2$ but $c\sim 2$ (which is allowed indeed), and then lowering of $m_0$, may solve this problem and can make $QDIH$ models safe.  

The discussion so far tells that on phenomenological ground, there is no dispute on the existence of quasi-degenerate neutrinos with $m_{i} < 0.1 eV$, in nature. But the question whether it is of NH type or IH type, is still not clear. In this regard, we expect that possible answer may emerge from the observed Baryon asymmetry of the universe $(\eta_B=6.5^{+0.4}_{-0.5}\times 10^{-10})$ \cite{Francis:2012jj, Francis:2012jk,Singh:2009kf,Sarma:2006xk}. The calculation of $\eta_{B}$ requires the texture of heavy right handed Majorana neutrino mass matrix, $M_{RR}$. With a suitable choice of  Dirac neutrino mass matrix, $m_{LR}$ allowed by $SO(10)$ GUT, we can transit from $m_{LL}^{\nu}$ (parametrized so far) to $M_{RR}$ by employing the inversion  of Type I see-saw formula,  $M_{RR}=-m_{LR}^{T} m_{LL}^{\nu^{-1}} m_{LR}$. We hope that significant physical insight can be fetched from this approach and it would be possible to figure out the most favorable QDN models out of the six. Unlike, in Refs.\,\cite{Francis:2012jj, Francis:2012jk}, the parametrization of mass matrices involves only two free parameters ($\alpha, \eta$) and no other input constant terms which are also different for TBM and TBM deviated scenarios. The prediction of $\theta_{12}$ involves $(\eta,\epsilon,c,d)$, whereas in our parametrization it depends on $\eta$ only. With minimum number of parameters, we have achieved a better control over mass matrices. In contrast to Refs.\,\cite{Francis:2012jj, Francis:2012jk}, with our parametrization of $m_{LL}^{\nu}$, certain analytical structure of $U_{\nu L}$ is also possible. We hope, this parametrization will be useful for other phenomenological studies also.


\end{document}